\documentclass[twocolumn,prb,aps,showpacs,amsmath,amssymb,superscriptaddress]{revtex4}
\usepackage{amsfonts}
\usepackage{bbm}
\usepackage{dcolumn}
\usepackage{bm}
\newcommand\mycite[1]{\raisebox{-0.5em}{\Large\cite{#1}}}
\usepackage[bookmarks=true,colorlinks=false,hyperindex=true,pdfstartview=FitH]{hyperref}
\usepackage{graphicx,subfigure}
\usepackage{epstopdf}
\begin{document}
\title{Effect of magnetic field on a magnetic topological insulator film with structural inversion asymmetry}
\author{Shu-feng Zhang}
\affiliation{Institute of Physics, Chinese Academy of Sciences, Beijing 100190, China}
\author{Hua Jiang}
\affiliation{Department of Physics and Jiangsu Key Laboratory of Thin Films, Soochow University, Suzhou 215006, China}
\author{X. C. Xie}
\affiliation{International Center for Quantum Materials, School of Physics, Peking University, Beijing 100871, China}
\affiliation{Collaborative Innovation Center of
Quantum Matter, Beijing 100871, China}
\author{Qing-feng Sun}
\email{sunqf@pku.edu.cn}
\affiliation{International Center for Quantum Materials, School of Physics, Peking University, Beijing 100871, China}
\affiliation{Collaborative Innovation Center of
Quantum Matter, Beijing 100871, China}

\begin{abstract}
The effect of magnetic field on an ultrathin magnetic topological
insulator film with structural inversion asymmetry is investigated.
We introduce the phase diagram, calculate the Landau-level spectrum
analytically and simulate the transport behavior in
Landauer-B\"uttiker formalism. The quantum anomalous Hall phase will
survive increasing magnetic field. Due to the two spin polarized
zero modes of Landau levels£¬ a nontrivial phase  similar with
quantum spin  Hall effect can be induced by magnetic field£¬ which
is protected by structural inversion symmetry. Some exotic
longitudinal and Hall resistance plateaus with fractional values are
also found in a six terminal Hall bar, arising from the coupling
between edge states due to inverted energy band and Landau levels.
\end{abstract}


\maketitle

\section{Introduction}

The quantum Hall effect (QHE),\cite{IQHE} as a quantized version of
Hall effect,\cite{Halleffect} occurs when dissipationless chiral
edge states form at sample edges due to the Landau level (LL)
quantization in the perpendicular magnetic field. QHE is also
suggested to exist without LLs and magnetic
field.\cite{QAH(Haldane),QAH(proposal),Yu(science)} In this type of
QHE, the chiral edge states are due to the inverted energy band
structure. It is considered the quantized version of the anomalous
Hall effect\cite{AHE_E} and named as quantum anomalous Hall (QAH)
effect. If time reversal symmetry (TRS) is preserved,
counterpropagate dissipationless edge states can form due to
inverted energy band structure,\cite{QSH_T} which is named quantum
spin Hall (QSH) effect as a quantized version of spin Hall
effect.\cite{SpinHall} In QSH effect, two counterpropagate spin
polarized edge states will vanish the Hall resistance while the
longitudinal resistance is quantized to be $h/(2e^2)$.\cite{QSH_E,jiang1} QSH effect is
soon discovered\cite{QSH_E} after the prediction.\cite{QSH_T}
However the QAH effect is observed very recently by Chang $et$
$al$\cite{Xue_QAH} though predicted much earlier.\cite{QAH(Haldane)}

In Ref.(\mycite{Xue_QAH}), QAH effect is observed in a magnetically
doped topological insulator (TI) film. The TI is a new phase of
matter which behaves as an insulator in the bulk but has
dissipationless edge states (two dimensional) or surface states
(three dimensional).\cite{review(Ti)} The low energy excitation is
the massless Dirac fermion for isolated surface states. But if the
TI film is thin enough, inter-surface tunnelling will occur
resulting in an energy gap to make the excitation
massive.\cite{NJP(Shan),PRB(Lu)} By magnetically doping the TI film,
ferromagnetic order will form resulting in an exchange field at low
temperature. If the exchange field overcomes the energy gap and is
perpendicular to the film, QAH effect will occur.\cite{Yu(science)}
With gates on both surfaces of the film, the potential of each
surface as well the Fermi energy of the sample can be tuned
conveniently. The potential difference between two surfaces acts as
a structure inversion asymmetry (SIA) term. In
Ref.(\mycite{Xue_QAH}) measurement under magnetic field is also
performed, but only the QAH resistivity plateau is observed even in a
strong magnetic field. The QHE keeps missing because of the low
mobility of the sample. Therefore it's still unknown how the QAH
effect and QHE will interplay and what's the effect of SIA potential in
presence of magnetic field.

Existing literature mainly focus on the LL spectrum in two
dimensional (2D) infinite TI
film.\cite{effective_H_Liu,QSH_E,LLTIfilm,PRB(L.Sheng),JAPLL} In a
perpendicular magnetic field, both the massless and massive Dirac
fermions will quantize into LLs.\cite{book_QED1965} The isolated
surface state accounts for the massless case\cite{effective_H_Liu}
and the $\bold k$-independent inter-surface tunneling acts as the
mass term which can be tuned by the thickness.\cite{LLTIfilm} Indeed
there has already been several experimental progress. The quantized
LLs for massless Dirac fermions have been measured in
$Bi_2Se_3$,\cite{DiracLL_TI_E} $Sb_2Te_3$,\cite{PRL(Y.P.Jiang)}
$Bi_2Te_3$,\cite{DLL(Okada)} and $HgTe$.\cite{DLL(HgTe)} However the
effect of $\bold k$-dependent inter-surface tunneling, SIA potential and
exchange field is still not studied theoretically. The $\bold
k$-dependent tunneling will make two zero modes of LLs linearly
depend on the magnetic field and cross to form a QSH-like phase. SIA
potential term is necessary to describe the potential distribution in the
perpendicular direction even for a film grown on a substrate with a
single back gate.\cite{lu2013QAH}

Few work focuses on the transport simulation to understand the
interplay between QAH effect and QHE. Though LL spectrum is helpful,
transport simulation is still necessary to give a complete
explanation. The reason is that edge state occurs at the boundary,
then only in a finite size sample the coupling between the edge
state due to inverted band structure and edge state due to LLs are
uncovered directly. Indeed, the coupling gives rise to exotic
behavior missing in LL spectrum.

In this paper we study the property of a magnetic TI film with SIA
in a perpendicular magnetic field. Both the LL spectrum and
transport simulation are investigated. The two zero modes are spin
polarized with opposite Chern number to form a QSH-like phase
protected by structural inversion symmetry at small magnetic field.
Magnetic field can't break the QAH phase with Chern number $C=1$.
When it comes to the case of nanoribbon with open boundary
condition, the edge state due to the inverted band structure will
couple with the edge state due to LLs at the same edge to form an
exotic energy band. Correspondingly exotic resistance plateaus with
nonzero longitudinal component can be observed in a six terminal
Hall bar. Besides we find that it's helpful to classify the LLs
according to the energy band they correspond to. The coupling
between edge state due to inverted energy band and edge states due
to LLs corresponding to trivial energy band is weaker than that to
nontrivial band.

The rest of the paper is organized as follows. In Sec. II, we
introduce the Hamiltonian and the phase diagram in absence of
magnetic field. In Sec. III, we derive the LL spectrum in the infinite
system analytically. In Sec. IV, we study the transport properties
of the six-terminal system in Landauer-B\"uttiker formalism. A
summary is given in the last section.

\begin{figure}
\setlength{\abovecaptionskip}{-0.05cm}
  \centering

  \subfigure{
    \includegraphics[width=0.994\columnwidth,height=0.39\columnwidth]{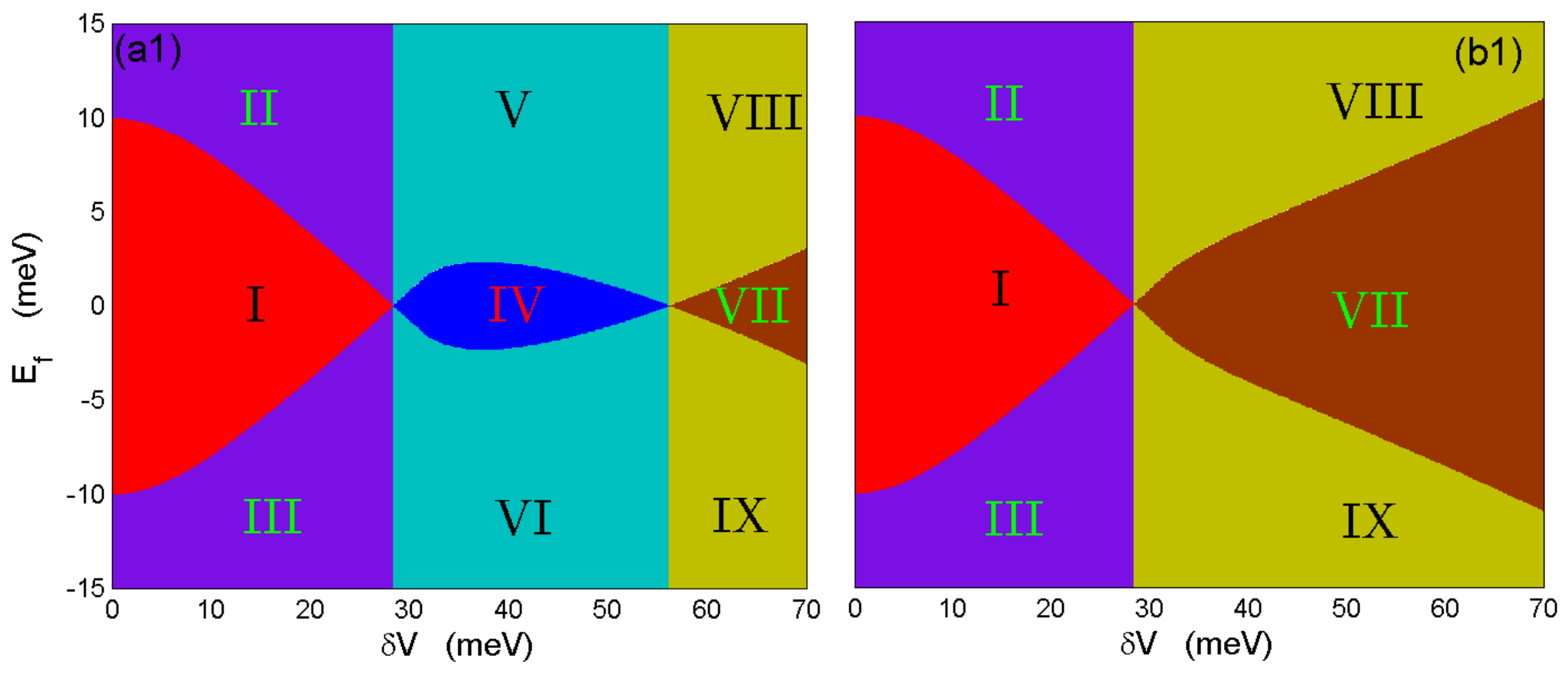}}
  \vspace{-0.55cm}

  \subfigure{
    \includegraphics[width=0.998\columnwidth,height=0.37\columnwidth]{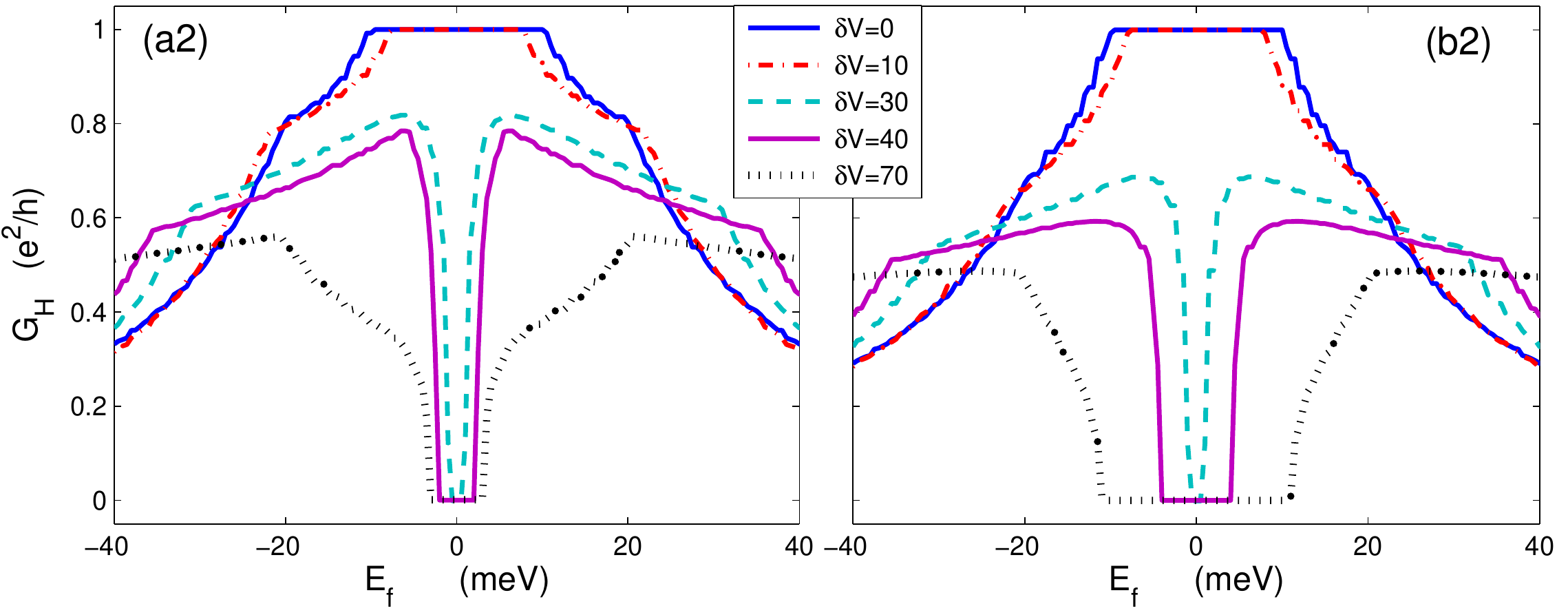}}
    \vspace{-0.55cm}

  \subfigure{
    \includegraphics[width=1\columnwidth,height=0.75\columnwidth]{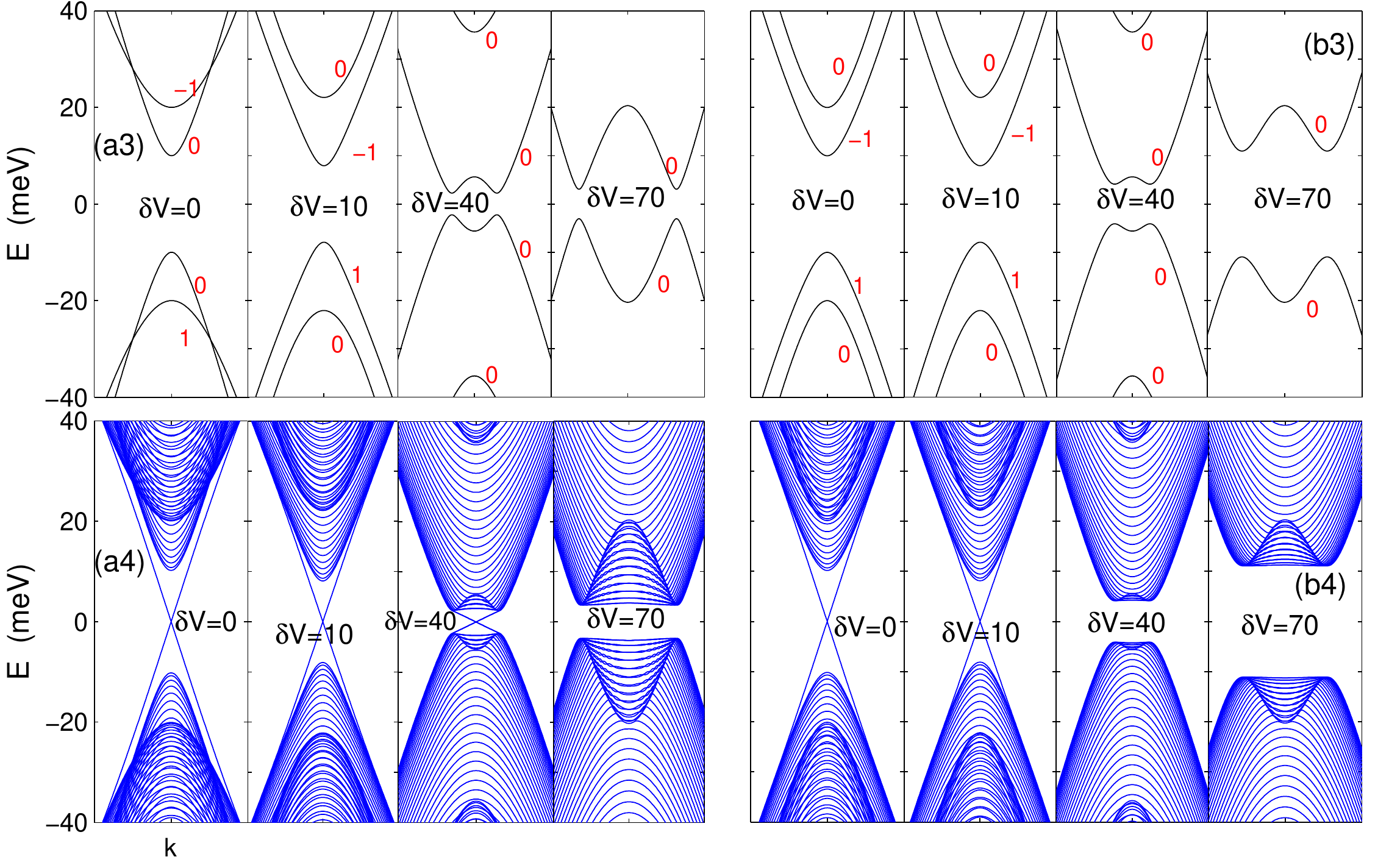}}
  \caption{(Color online) (a1,b1) phase diagrams in the plane of SIA potential $\delta V$ and Fermi energy $E_f$.
Regions I-VI are in inverted regime. Region I (IV) is the QAH (QPH)
phase with Fermi level in the bulk gap, and  II,V (III,VI) are the
n(p)-doped inverted system. Region VII is the normal insulator phase
with Fermi level in the bulk gap. Regions VIII and IX are in normal metal phase. (a2,b2) Hall conductance as a function of Fermi
energy for several $\delta V$. (a3,b3) the energy dispersion of
the 2D system in a certain direction. (a4,b4) the energy dispersion of a
nanoribbon with the ribbon width $480nm$. (a1-4) for $m_0/B<0$ and
(b1-4) for $m_0/B>0$ with the exchange field $M=-15meV$. Energy in
unit of $meV$ in this paper if not indicated.} \label{figphase}
\end{figure}

\section{Hamiltonian and Phase diagram}
\label{sec_phase}
The effective Hamiltonian near the $\Gamma$ point for bulk $Bi_2Se_3$ is given by\cite{HJZhang}
\begin{eqnarray}
  H_{bulk}\!\!=\!\!\epsilon_0({\bold k})I_{4\times4}\!\!+\!\!\left(\begin{array}{cccc}
  M({\bold k})&-iA_1 \partial_z&0&A_2 k_-\\-iA_1 \partial_z&-M({\bold k})&A_2 k_-&0\\0&A_2k_+&M({\bold k})&iA_1 \partial_z\\A_2k_+&0&iA_1 \partial_z&-M({\bold k})
  \end{array}\right),  \nonumber\\
\end{eqnarray}
in the basis formed by the hybridized states of $Se$ and $Bi$ $p_z$ orbitals, denoted as $(|p1_z^+,\!\uparrow\rangle,|p2_z^-,\!\uparrow\rangle,|p1_z^+,\!\downarrow\rangle,|p2_z^-,\!\downarrow\rangle)^T$.
where $k_{\pm}= k_x \pm ik_y$, $\epsilon_0({\bold k}) = C - D_1 \partial_z^2 + D_2 k^2$,
$M({\bold k})=M+B_1\partial_z^2-B_2k^2$,
and $k^2 = k_x^2+ k_y^2$,
with $A_1$, $A_2$, $B_1$, $B_2$, $C$, $D_1$, $D_2$ and $M$ being the model parameters.

For a thin film with surfaces perpendicular to the $z$-direction, surface states will emerge with effective Hamiltonian\cite{Yu(science),note1}
\begin{eqnarray}
H_{sf}=\tilde A_2 (k_y \sigma_x\!-\!k_x\sigma_y)\tau_z+m_k \tau_x,
\end{eqnarray}
in basis $(|t\!\uparrow\rangle,\,|t\!\downarrow\rangle,|b\!\uparrow\rangle,\,|b\!\downarrow\rangle)^T$,
where $t\,\, (b)$ represents the top (bottom) surface states and
$\uparrow(\downarrow)$ represent the up(down)-spin states.
$|\!\!\uparrow\!\!(\downarrow)\!\rangle$ is the superposition of the up(down)-spin hybridized states of $Se$ and $Bi$ $p_z$ orbitals.
$\sigma_{x,y,z}$ and $\tau_{x,y,z}$ are Pauli matrices acting on spin
space and mixing the top and bottom surface states respectively.
The first term in Hamiltonian describes the free surface states with
$\tilde A_2=\hbar v_F$ describing the Fermi velocity of surface state. The second term is
the coupling between the top and bottom surface states with
$m_k=m_0+B k^2$. Inter-surface tunneling will open a gap
with width $2m_0$, then the ultrathin three dimensional TI film is
in QSH phase with two inverted band structure for $m_0/B<0$ and
topological trivial phase for $m_0/B>0$ depending on film
thickness.\cite{Yu(science),QSH_NI_Osci(Liu),NJP(Shan),PRB(L.Sheng2012)}

Then the Hamiltonian of the thin magnetic TI film with electric potential on both surfaces can be written
as,
\begin{eqnarray}\label{h1}
 H\!=H_{sf}+\!M\sigma_z \!+\!\frac{\delta V}{2} \tau_z
\end{eqnarray}
The second term is due to the exchange field along
$z$ direction to describe the ferromagnetic order induced by
magnetic dopants. It acts just like the Zeeman energy. A moderate
enough exchange field ($|M|>|m_0|$) will alter the energy band
resulting in only one inverted band to give rise to the QAH phase
both for $m_0/B<0$ and $m_0/B>0$ . The last term is
the SIA potential, i.e. it describes the effect of the gate voltages on
top and bottom surfaces with magnitudes $\pm\delta V/2$ respectively.

There is only a difference of unitary transformation between
$H(-\delta V,m_k,v_F,M)$, $H(\delta V,-m_k,v_F,M)$, $H(\delta
V,m_k,-v_F,M)$, $-H(\delta V,m_k,v_F,-M)$ and $H(\delta
V,m_k,v_F,M)$. Therefore we only consider the positive SIA potential
$\delta V$, the positive $m_0$ and the negative $M$. Accordingly
parameters in Hamiltonian are chosen as $v_F=4.53\times10^5ms^{-1}$,
$m_0=5meV$ and $B=\pm800meV\cdot nm^2$ in this
paper.\cite{nphys(Y.Zhang)} Both cases $m_0/B<0$ and $m_0/B>0$
are studied respectively.
For the case $M=0$ the system lies in QSH phase which has been studied in detail and well understood.\cite{review(Ti),QSH_E,PRB_HgTe(J.C.Chen),jiang2}
Therefore we mainly focus on the case $|M|>m_0$ and set $M=-15meV$ so that the QAH phase will emerge for both $m_0/B<0$ and $m_0/B>0$ in absence of SIA potential ($\delta V=0$).

The continuous energy band of the Hamiltonian (\ref{h1}) is derived
as,
\begin{eqnarray}\label{eqdis}
  \epsilon^2&=&\hbar^2v_F^2k^2+m_k^2+M^2+\delta V^2/4\nonumber\\
&\pm&2\sqrt{m_k^2 M^2+M^2\delta V^2/4+\hbar^2v_F^2k^2\delta V^2/4}.
\end{eqnarray}
An energy gap can be opened by the inter-surface tunneling $m_k$,
exchange field $M$ or SIA potential $\delta V$. There are two branches
split by exchange field and SIA potential in both electron-like and
hole-like energy bands. Fig.\ref{figphase} (a3,b3) show energy
dispersion curves. The two branches will cross for $m_0/B<0$ and
$\delta V=0$. However, even a tiny $\delta V$ can break the
crossing. In the following we will show that this crossing leads to
interesting phenomena.

Tuning the potential difference $\delta V$ adiabatically, the energy
gap will close and reopen according to Eq.(\ref{eqdis}). For
$m_0/B<0$ the gap closing occurs twice for $|\delta
V|/2=\sqrt{M^2-m_0^2}$ at $\Gamma$ point and $|\delta
V|/2=\sqrt{M^2+ \hbar^2v_F^2|m_0/B|}$ at
$|\bold{k}|\!=\!\sqrt{-m_0/B}$ in $\bold{k}$-space as shown in
Fig.\ref{figphase}(a1). For the $m_0/B<0$ case, the gap only closes
and reopens once for $|\delta V|/2=\sqrt{M^2-m_0^2}$ at $\Gamma$
point. Fig.\ref{figphase}(a/b1) show the phase diagram in the plane
of the SIA potential $\delta V$ and Fermi energy $E_f$, in which
nine regions emerge for the $m_0/B<0$ case and six regions for the
$m_0/B>0$ case.
To determine the topological property of each region we calculate
the Chern number which equals the zero-temperature Hall conductance
by $G_H\!=\!C \frac{e^2}{h}$ and can be calculated via the Kubo
formula\cite{KuboChern,note2}
\begin{eqnarray}\label{ChernNumber}
G_{H}\!\!=\!\!\frac{2e^2}{\hbar}\int_{BZ} \frac{dk_xdk_y}{(2\pi)^2}\!\!\sum_{\epsilon_l<E_f<\epsilon_n }\!\!\!
Im\frac{\langle l|\frac{\partial H}{\partial k_x} |n\rangle \langle n|\frac{\partial H}{\partial k_y}|l\rangle}{(\epsilon_l-\epsilon_n)^2},
\end{eqnarray}
where BZ represents first Brillouim zone, and $\epsilon_{l/n}$ and
$|l/n\rangle$ are the corresponding eigenenergy and eigenstate
respectively. Results shown in Fig.\ref{figphase}(a2,b2) indicate
that regions I-III in Fig.\ref{figphase}(a1,b1) are in inverted
regime. To determine the property of regions IV-IX, we calculate the
energy dispersion of a nanoribbon with open boundary condition along
$y$ direction to check directly whether an edge state exists or not.
Results in Fig.\ref{figphase}(a4,b4) suggest that regions IV-VI and
VII-IX are in inverted regime and normal insulator regime
respectively. And region IV is a quantum pseudo-spin (QPH)
phase\cite{NJP(Shan),PRB(L.Sheng2012)} with two counterpropagating
edge states at the same edge of the sample. In short it goes through
QAH phase, QPH phase and normal insulator phase as $\delta V $
increases from zero for $m_0/B<0$. In contrast, in the case
$m_0/B>0$, it directly goes from QAH phase to the normal insulator
phase, and the QPH phase doesn't appear.

Integers in Fig.\ref{figphase}(a2,b2) show the Chern number of the
corresponding band. From Fig.\ref{figphase}(a4,b4) and the Chern
number, we indicate that the inner (outer) two bands are trivial
(nontrivial) for $m_0/B<0$ but nontrivial (trivial) for $m_0/B >0$
in absence of SIA potential ($\delta V=0$). But even for a small
nonzero SIA potential $\delta V$, the two inner bands are nontrivial
and the outer ones are trivial in the inverted regime. While $\delta
V$ is large enough ($\delta V/2>\sqrt{M^2-m_0^2}$ for $m_0/B>0$ or $\delta
V/2>\sqrt{M^2+\hbar^2 v^2_F |m_0/B|}$ for $m_0/B<0$), the system is in the
normal regime, and then all four bands are trivial.

\section{magnetic field and Landau levels}
\label{secLLs}

In this section the LL spectrum of the 2D infinite system is
studied. It is useful for understanding magnetic properties, such as
the Hall effect, magneto-optics and Shubnikov-de Haas oscillation.

Magnetic fields will induce two types of contributions, the orbital
effect and the Zeeman effect. The Zeeman term takes the same form as
the exchange field but with a lesser order of magnitude. Therefore
the Zeeman term can be neglected in the effective Hamiltonian. The
orbital effect can be included by Peierls substitution $\bold {k}
\to \bold\pi= \bold k+\frac{e}{\hbar}\bold A $ in which $\bold
A=(B_z y,0,0)$ for magnetic field $(0,0,-B_z)$ and $e$ is the
magnitude of electron charge. To calculate the LL spectrum we define
the annihilation and creation operator for the harmonic oscillator
function $\phi_N$ as $\hat a=l_c \pi_+ /\sqrt{2},\,\hat
a^\dag=l_c\pi_-/\sqrt{2}$ satisfying $[\hat a, \hat a^\dag]=1$,
$\hat a\, \phi_N=\sqrt{N} \phi_{N-1}$, $ \hat
a^\dag\,\phi_N=\sqrt{N+1}\phi_{N+1}$ with
$l_c=\sqrt{\frac{\hbar}{eB_z}}$, $\pi_{\pm}=\pi_x\pm i\pi_y$ and
$N=1,2,3,...$ .\cite{PRL(2004)(S.Q.Shen)(SR)} Then the Hamiltonian
can be written as
\begin{equation}
\tilde H\!=\!\!\left(\!\begin{array} {cccc}
 M+\delta V/2&i\omega_1 a^\dag&m_N-\alpha/2&0\\-i\omega_1a&-M+\delta V/2&0&m_N-\alpha/2\\ m_N-\alpha/2&0&M-\delta V/2&-i\omega_1 a^\dag\\0&m_N-\alpha/2&i\omega_1 a&-M-\delta V/2
\end{array}\!\right)\nonumber\\
\end{equation}
in which $m_N=m_0-\alpha\hat a^\dag\hat a$,
$\omega_1=v_F\sqrt{2eB_z\hbar}$ and $ \alpha=-\frac{2eB_zB}{\hbar}$.
With the wave function ansatz $\psi_N=(f_1^N \phi_{N},f_2^N
\phi_{N-1},f_3^N \phi_{N},f_4^N \phi_{N-1})^T$, the Hamiltonian can
be written as
\begin{eqnarray}
\tilde H^\prime\!\!=\!\!\left(\begin{array} {cccc}
 M+\delta V/2&i\omega_N&\tilde m_N-\alpha/2&0\\-i\omega_N&-M+\delta V/2&0&\tilde m_N+\alpha/2\\ \tilde m_N-\alpha/2&0&M-\delta V/2&-i\omega_N\\0&\tilde m_N+\alpha/2&i\omega_N&-M-\delta V/2
\end{array}\!\!\right)\nonumber\!\!\!\!\!\!\!\\
\end{eqnarray}
in which $\tilde m_N=m_0-\alpha N$ and $\omega_N=v_F\sqrt{2e\hbar B_z
N}$. Then LLs can be solved from the secular equation
straightforwardly.

The zero mode of LLs can be derived as
$E_{0\pm}=M\pm\sqrt{(m_0-\alpha/2)^2+(\delta V/2)^2}$ with wave
function $\psi_0=(f_1^0 \phi_{0},0,f_3^0\phi_{0},0)^T$. In absence
of SIA potential, the two branches are defined as
$E_{0\pm}=M\pm(m_0+\frac{eBB_z}{\hbar})$ which depends linearly on
magnetic field. However the general analytical expression for
nonzero modes is tediously long. Here we only analytically give the
solutions in which case the electron-like and hole-like nonzero
modes are symmetrical:

\noindent case I: $B=0$, with only lowest-order inter-surface tunnelling,
\begin{eqnarray}
  E_N^2=&&\omega_N^2+(\frac{\delta V}{2})^2+M^2+m_0^2 \nonumber \\
\pm&&2\sqrt{\omega_N^2(\frac{\delta V}{2})^2+(\frac{\delta V}{2})^2M^2+M^2m_0^2},
\label{eqLL1}
\end{eqnarray}
\noindent case II: $M=0$, undoped TI film,
\begin{eqnarray}
E_N^2=&&\omega_N^2+(\frac{\delta V}{2})^2+(\frac{\alpha}{2})^2+\tilde m_N^2\nonumber\\
\pm&&2\sqrt{\omega_N^2(\frac{\delta V}{2})^2+(\frac{\delta V}{2})^2(\frac{\alpha}{2})^2+(\frac{\alpha}{2})^2\tilde m_N^2}.
\label{eqLL2}
\end{eqnarray}

\begin{figure}[htbp]
\centering
\includegraphics[width=1.0\columnwidth,height=1.5\columnwidth]{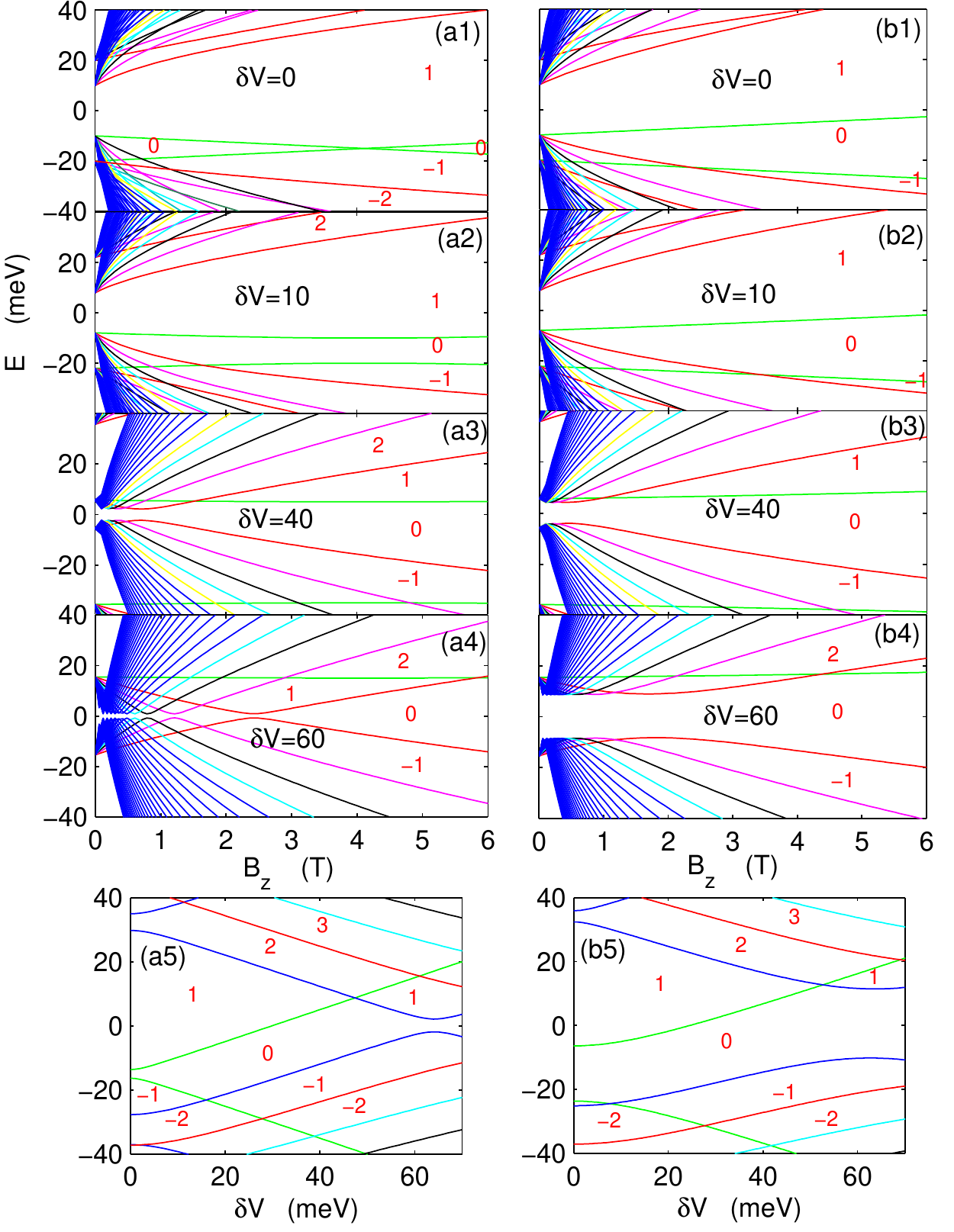}\\
\caption{(Color online) Landau level spectrum. (a/b1-4) LLs as a function of magnetic field for several $\delta V$.
(a/b5) LLs as a function of $\delta V$ for $B_z=3T$.
 Light green lines are the zero modes $E_{0\pm}$.
 $E_{0-}$ is the one lies below at small magnetic field.
 Red integers indicate the Chern number.
 (a1-5) for $m_0/B<0$, (b1-5) for $m_0/B>0$. The exchange field $M=-15meV$.
}
\label{figLLs}
\end{figure}

By eliminating the extra terms, i.e. inter-surface tunneling $m_k$,
exchange field $M$ and SIA potential $\delta V$, the LL spectrum of Dirac
electrons with linear energy dispersion is reduced into
$E_N=sign(N)v_F\sqrt{2eB_z\hbar |N|}$ with $N=0,\pm1,\pm2...$ which
is the well known result.\cite{book_QED1965,DiracLL_graphene_E}
However, the existence of these terms makes the dependence on $N$
and $B_z$ complex. Exchange field will break the symmetry of
electron-like and hole-like LLs. LLs will emerge and split to
realize fascinating phases by varying the magnetic field and SIA
potential as is shown in Fig.\ref{figLLs}.

\begin{figure*}
  \centering
  \subfigure{
    \includegraphics[width=1.6\columnwidth]{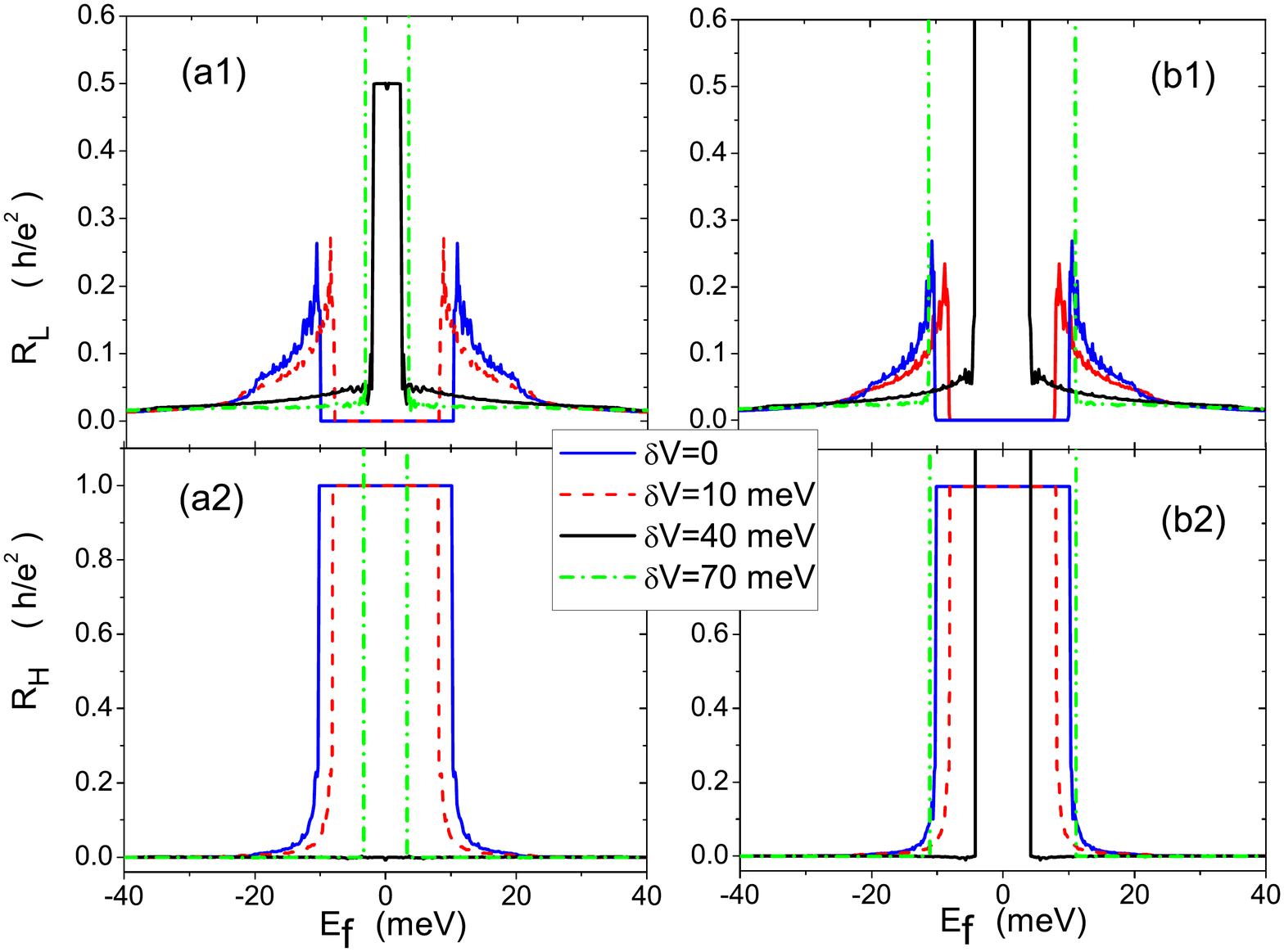}}
    \vspace{-10cm}

    \hspace{0.55cm}\subfigure{
    \includegraphics[width=0.5\columnwidth]{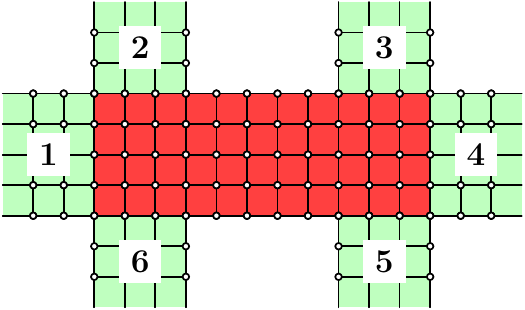}}

    \vspace{7cm}
\caption{(Color online) Hall resistance $R_H$ and longitudinal resistance $R_L$ as a
function of Fermi energy $E_f$ for several SIA terms in absence of
magnetic field. (a1-2) for $m_0/B<0$, (b1-2) for $m_0/B>0$. Inset is
the schematic of the Hall bar. The width of both the central region and six contacts is $480nm$.
  }
  \label{figB0}
\end{figure*}

If magnetic field is close to zero, $B_z\approx 0$, the nonzero LLs
will converge at $E_N\approx\pm (M\pm\sqrt{m_0^2+\delta V^2/4})$
while zero modes at $E_{0\pm}\approx M\pm\sqrt{m_0^2+\delta V^2/4}$.
Accordingly LLs can be classified into four groups as is shown in
Fig.\ref{figLLs} (a/b1-4). Each group corresponds to one branch of
the four energy bands. For instance, the group that converges at
$M-\sqrt{m_0^2+\delta V^2/4}$ corresponds to the negative outer
energy band. Red integers in Fig.\ref{figLLs} represent Chern
numbers of each region from which the Chern number of each mode of
LLs can be indicated. The two zero modes have opposite Chern
numbers. If $E_{0-}$ corresponds to a trivial band it will be
hole-like with negative Chern number $C=-1$ the same as nonzero
modes in the same group, see Fig.\ref{figLLs}(a2-4,b1-4). However
$E_{0-}$ will be electron-like with positive Chern number different
from other modes in the same group if it corresponds to a nontrivial
band, see Fig.\ref{figLLs}(a1).

In Fig.\ref{figLLs}(a1) $E_{0-}$ lies below $E_{0+}$ and cross with
the hole-like inner LLs with negative Chern number for
$B_z\!<\!B_c=-\frac{\hbar m_0}{e B}\approx4.1T$. Here $4.1T$ also is
the crosspoint of $E_{0+}$ and $E_{0-}$ (see the light green lines
in Fig.\ref{figLLs}(a1)). As a result, in the region encircled by
$E_{0-}$ and $E_{0+}$, there will be one anticlockwise edge state,
$E_{0-}$, and several clockwise edge states for a finite size sample
with appropriate Fermi energy $E_{0-}\!\!<\!E_f\!<\!E_{0+}$.
Decreasing the magnetic field from $B_c$, more clockwise edge states
will emerge. In the region $E_{0-}<E_f<E_1$ and $0<B_z<0.7T$ ($0.7T$
is the crosspoint of $E_{0-}$ and $E_1$, the $N=1$ mode of inner hole-like
LLs), there are more than one clockwise edge states but the
anticlockwise edge state is one always.

The regime with vanished Chern number in Fig.\ref{figLLs}(a1) is
similar with that in inverted HgTe/CdTe quantum well in
Ref.(\mycite{QSH_E,PRB_HgTe(J.C.Chen)}). Two zero modes are up-spin
polarized completely with opposite Chern number. Therefore zero
modes lead to a QSH-like phase with vanished Chern number and
counterpropagating edge states for $B_z<B_c$. By increasing the
magnetic field from $B_c$, two zero modes will cross with each other
to form a normal-insulator gap. It's well known that magnetic field is to break the QSH phase,\cite{PRB_HgTe(J.C.Chen)}
here we show that the moderate magnetic field can result in and large magnetic field can break the
QSH-like phase. Besides, since
$E_{0+}-E_{0-}=2\sqrt{(m_0-\alpha/2)^2+(\delta V/2)^2}$, the SIA
term will break the crossing of zero modes (see
Fig.\ref{figLLs}(a1-2)), leading to a normal-insulator gap though
they are in the same phase for $|\delta V|/2<\sqrt{M^2-m_0^2}$ in
absence of magnetic field. The reason is that the presence of SIA
potential makes the topological properties of inner bands and outer bands
exchange abruptly. Exchange field will not contribute since it can
not lead that change. For $m_0/B>0$, there is no cross between two
zero modes $E_{0\pm}$ (see Fig.\ref{figLLs}(b1-4)) and the QSH-like
phase does not emerge. In brief the QSH-like phase is
irrelevant with the exchange field and protected by structural
inversion symmetry and $m_0/B<0$.

In Fig.\ref{figLLs}(a/b1-2) there is only one region with Chern
number $C=1$ due to one zero mode for $|\delta
V|/2<\sqrt{M^2-m_0^2}$, which is just the boundary of QAH phase for
vanishing magnetic field. And for $B_z\approx0$ the width of this
region is given as $M+\sqrt{m_0^2+\delta
V^2/4}<E_f<-M-\sqrt{m_0^2+\delta V^2/4}$, which is just the
nontrivial gap of the QAH phase. Therefore the $C=1$ region can be
seen as a QAH phase. In other words QAH phase can survive by
increasing magnetic field consistent with the experiment result by
Chang $et$ $al$. \cite{Xue_QAH} Increasing $\delta V$ from zero, the
inner two groups of LLs corresponding to $\pm(M+\sqrt{m_0^2+\delta
V^2/4})$ will move upward or downward respectively to meet at
$B_z=0$ when $\delta V/2=\sqrt{M^2-m_0^2}$. Increase $\delta V$
anymore, two groups of LLs will exchange Chern number with a
insulator-gap formed, to lead a region with Chern number $C=0$.
Increasing $\delta V$ anymore, the gap will enlarge for $m_0/B>0$,
see Fig.\ref{figLLs}(b3-4). But it will close and reopen at about
$\delta V/2=\sqrt{M^2-m_0 \hbar^2v_F^2/B}$ for $m_0/B<0$, see
Fig.\ref{figLLs}(a3-4). To further show the effect of SIA potential, we
plot LLs against $\delta V$ at fixed magnetic filed $B_z=3T$ in
Fig.\ref{figLLs} (a5,b5). As expected the region with Chern number
$C=1$ at small $\delta V$ corresponds to the QAH phase, large SIA
potential will break the QAH phase.

\begin{figure*}
\subfigure{\includegraphics [width=1.6\columnwidth,height=0.8\columnwidth]{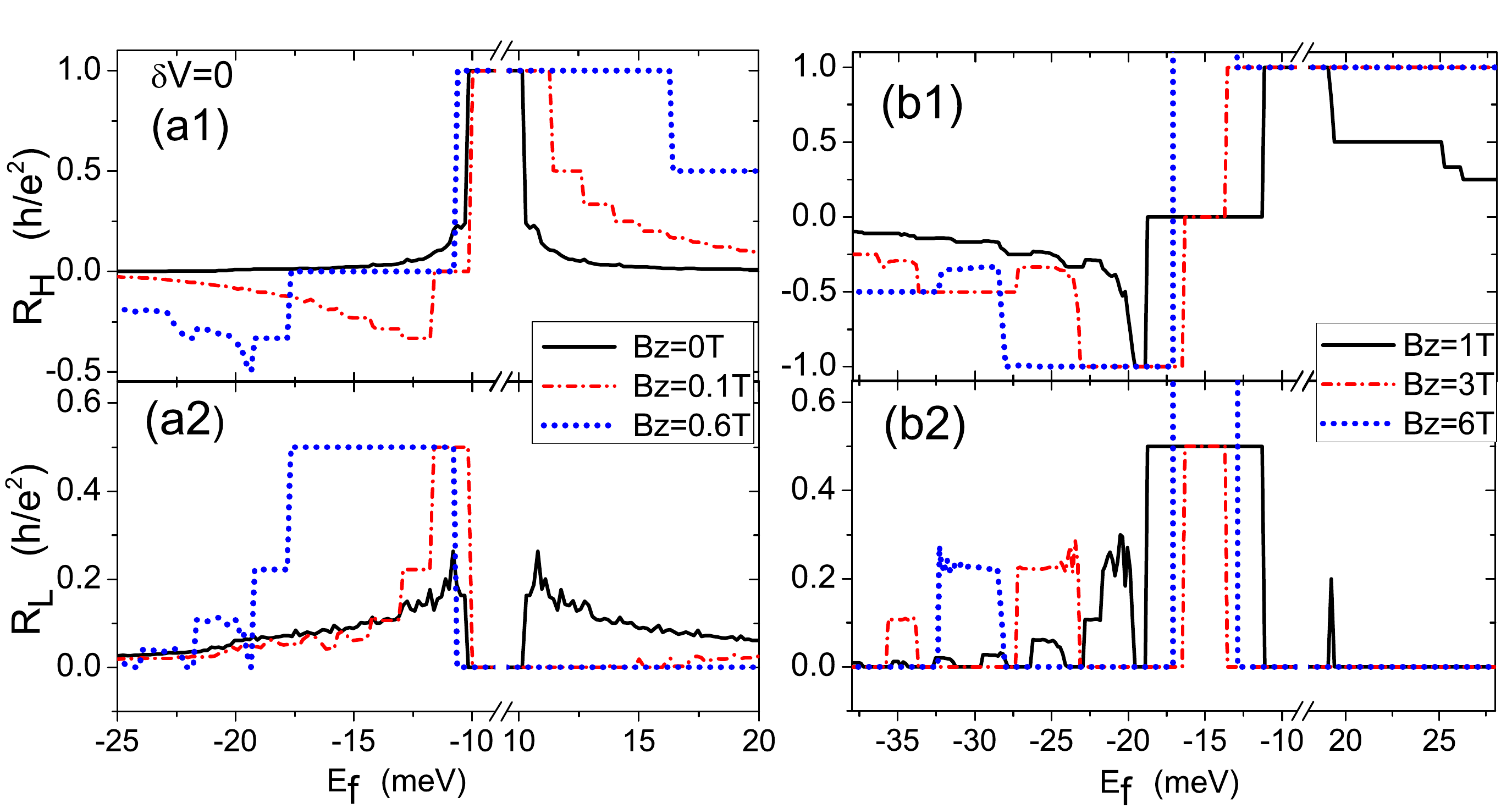}}
 \caption{(Color online) Hall resistance $R_H$ and longitudinal resistance $R_L$ as a function of Fermi energy with fixed magnetic field $B_z$. $\delta V=0$, $m_0/B<0$.}
  \label{figV0}
\end{figure*}

\begin{figure*}[htb]
  \centering
\hspace{-0.5cm}\subfigure{
    \includegraphics[width=0.75\columnwidth,height=.55\columnwidth]{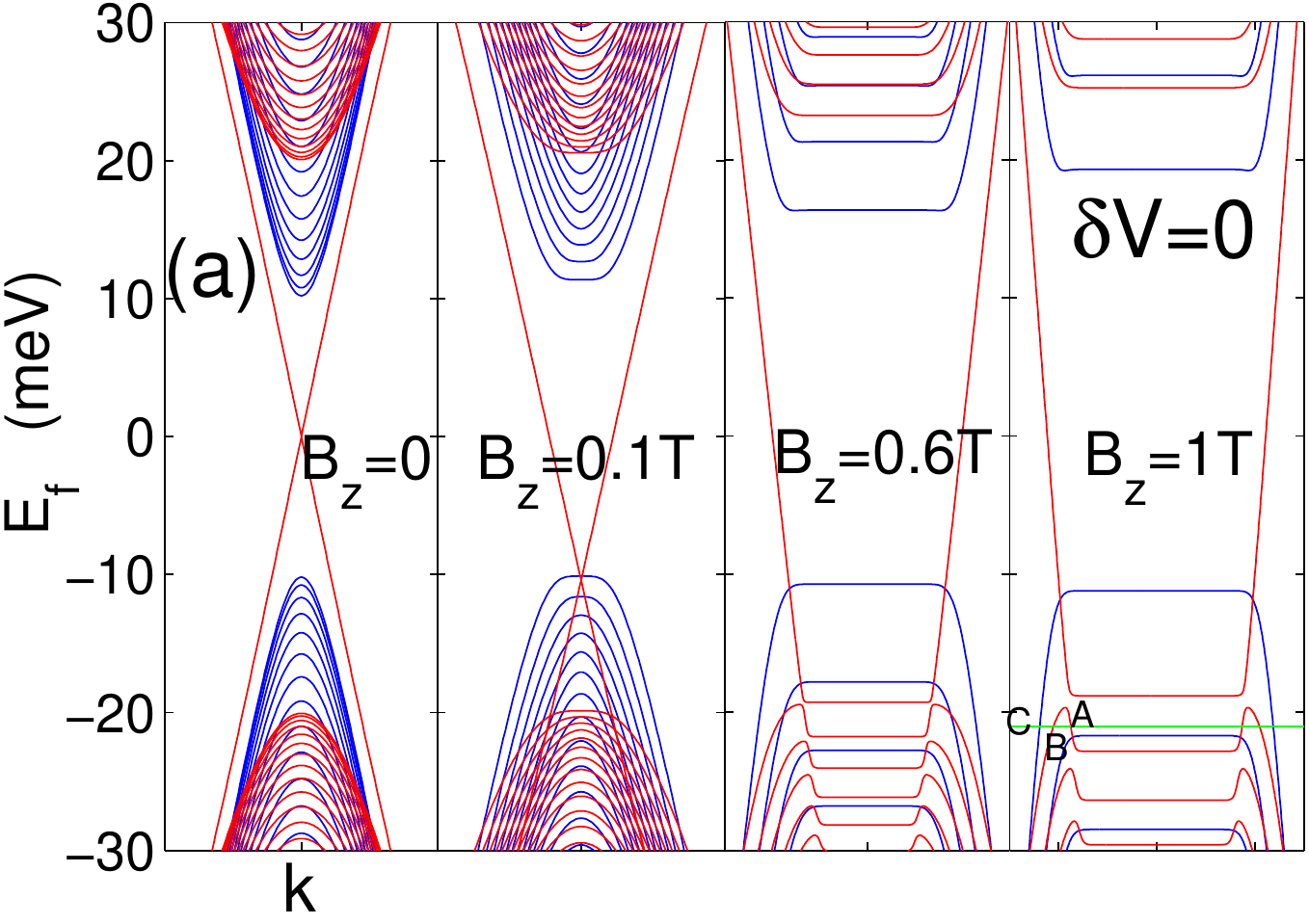}}
    \subfigure{
    \includegraphics[width=0.75\columnwidth,height=.55\columnwidth]{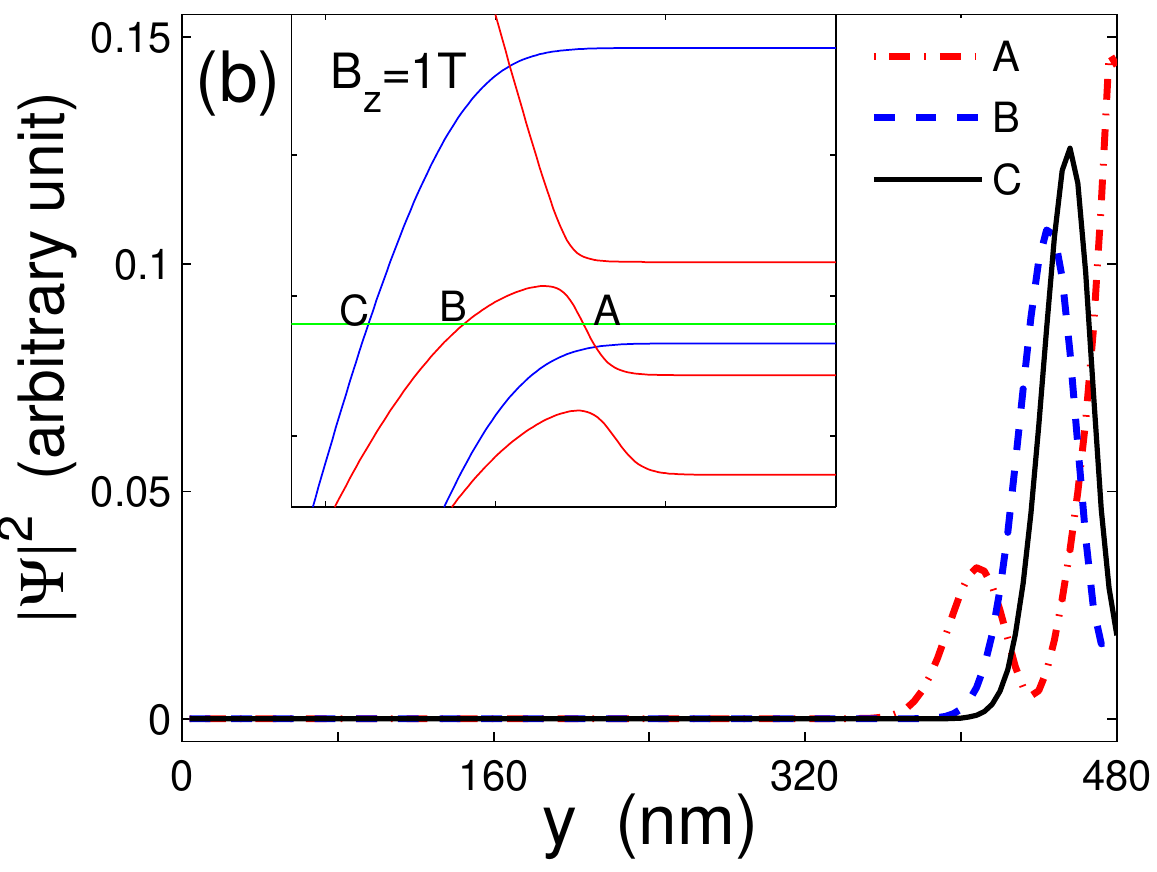}}
\caption{(Color online) (a) The energy dispersion for a 1D nanoribbon with width of 480 nm for several magnetic field respectively. (b) Wave function distribution along $y$ direction for the exotic edge states, (A,B) and edge state due to LL, C. Inset of (b) is the amplification of (a).
Parameters are identical to those in Fig.\ref{figV0}.}
  \label{figebV0}
\end{figure*}

\section{Transport analysis}
\label{secTransport}

This section is to study the transport properties of a standard Hall
bar with six leads (as shown in the inset of Fig.\ref{figB0}) by
Landauer-B\"uttiker formalism at zero temperature. On one hand
numerical results here will verify the analysis in
Sec.\ref{sec_phase} and Sec.\ref{secLLs}. On the other hand finite
size will lead to an intrinsic edge state due to inverted energy
band in the nontrivial energy gap. And in presence of magnetic
field, this intrinsic edge state will coexist or couple with the
magnetic edge state due to LLs leading to exotic transport
phenomena.

\begin{figure*}
\subfigure{\includegraphics[width=1.6\columnwidth,height=0.8\columnwidth]{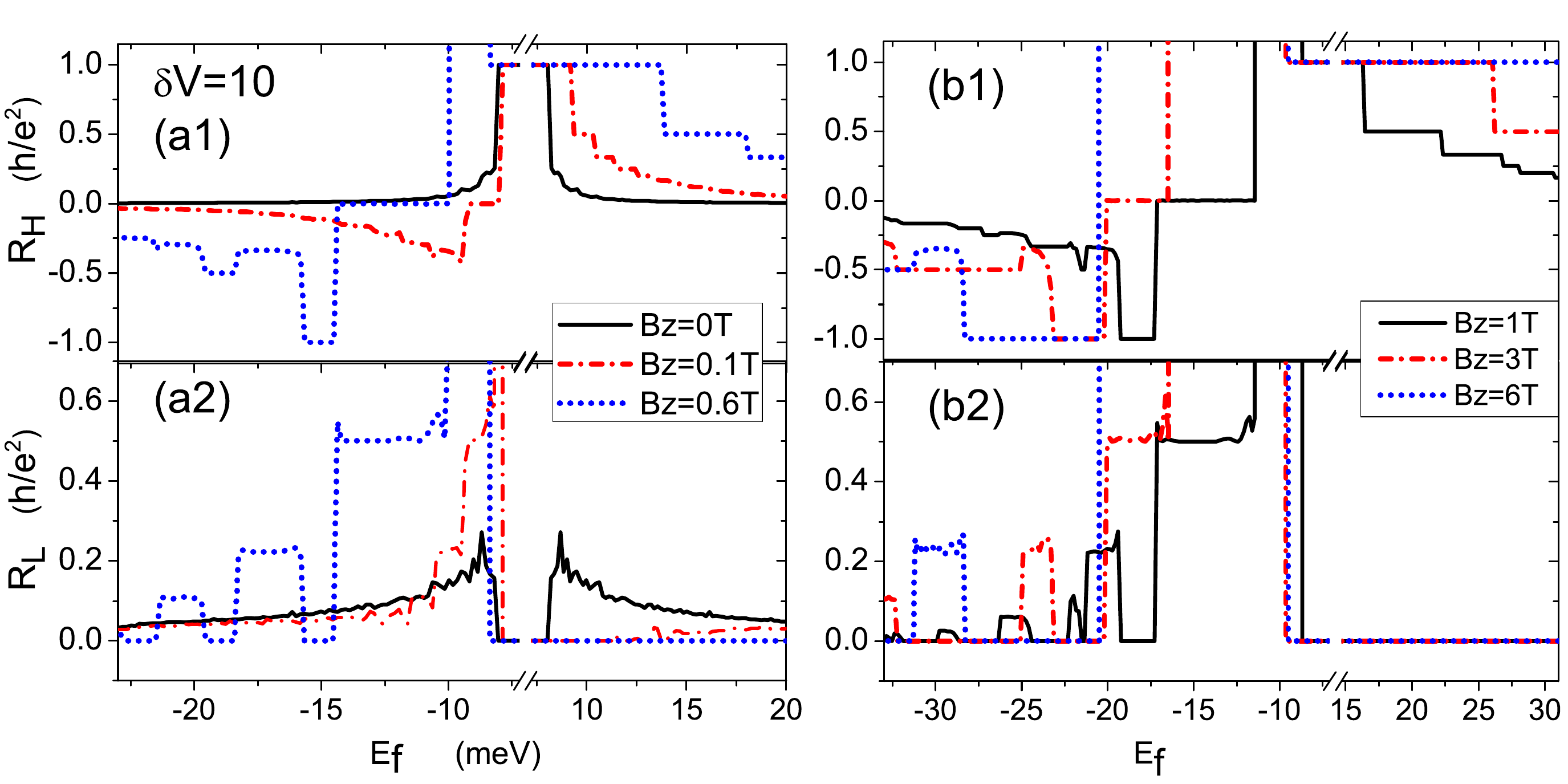}}%
 \caption{(Color online)  (a,b) Hall resistance $R_H$ and longitudinal resistance $R_L$ as a function of Fermi energy with fixed magnetic field $B_z$.
  $\delta V=10meV$, $m_0/B<0$.
}
  \label{figV10}
\end{figure*}

\begin{figure}
  \centering
\subfigure{
\includegraphics[width=.95\columnwidth]{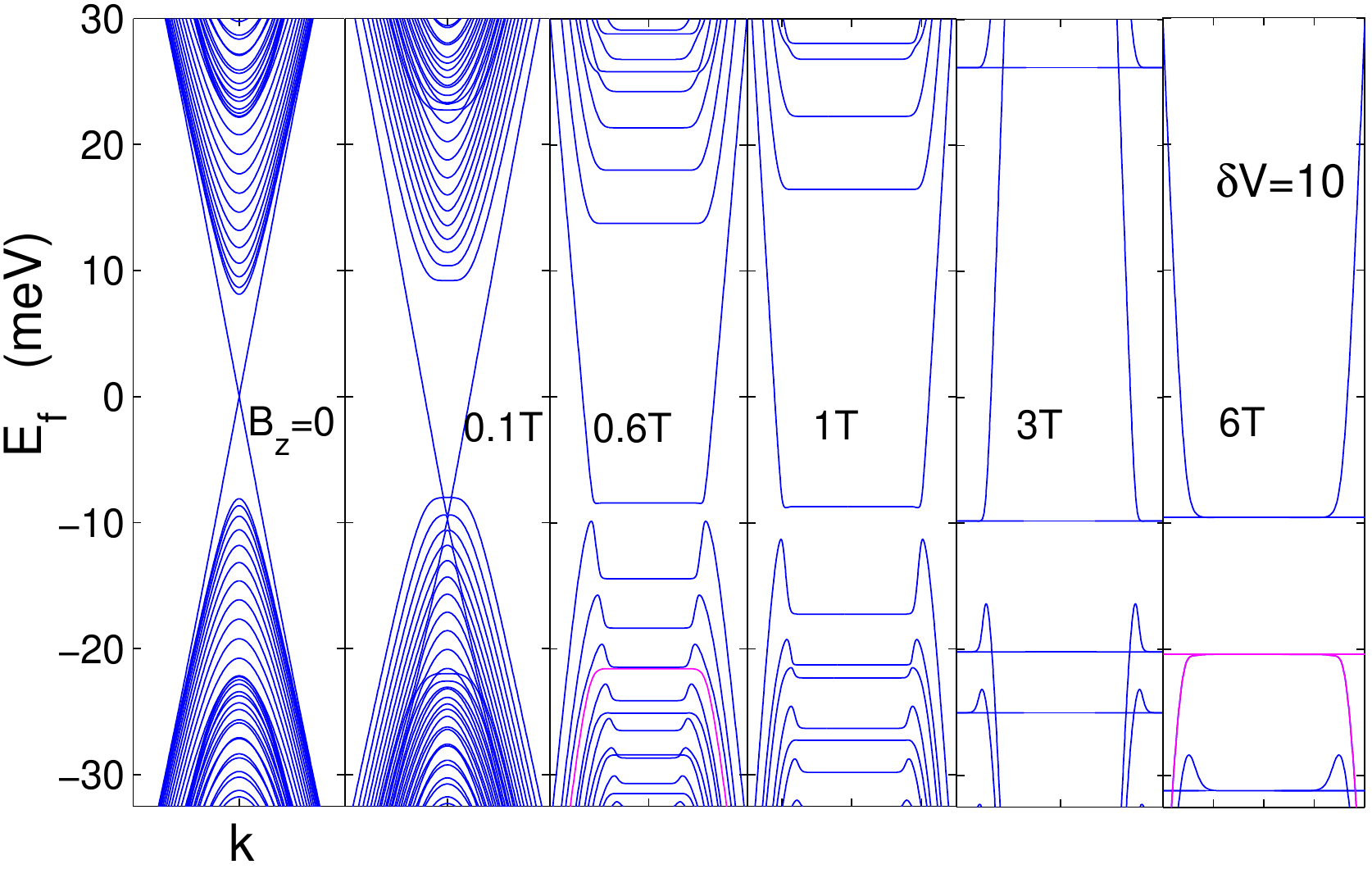}}

\caption{(Color online) The energy dispersion for a nanoribbon with width of 480 nm for several magnetic field. Parameters are identical to those in Fig.\ref{figV10}.
  }
  \label{figebV10}
\end{figure}

For convenience, we discretize the Hamiltonian in Eq.(\ref{h1}) in
square lattice:
\begin{eqnarray}\label{h2}
  H_d&=&\sum_{\bold x}[ \psi^\dag_{\bold x}T_0\psi_{\bold x}+(\psi^\dag_{\bold x}T_x\psi_{\bold x+\delta \hat x} +\psi^\dag_{\bold x}T_y\psi_{\bold x+\delta \hat y}+H.C.)]\nonumber\\
  T_0&=&M\sigma_z+\tilde m_0 \tau_x+\frac{\delta V}{2}\tau_z \nonumber\\
 T_x&=&(it_a\sigma_y\tau_z-B_a\tau_x)e^{i\phi_{{\bold x, \bold x + \delta \hat  x}}}\nonumber\\ T_y&=&(-it_a\sigma_x\tau_z-B_a\tau_x)e^{i\phi_{{\bold x, \bold x + \delta \hat  y}}}
\end{eqnarray}
where $\bold x$ is the site index and $\delta\hat x \,(\delta\hat
y)$ is the unit vector along $x\,(y)$ direction. $\psi_{\bold
x}=(a_{\bold x},b_{\bold x},c_{\bold x},d_{\bold x})^T$ represents
the four operators annihilate an electron on site ${\bold x}$ in
states $|t\uparrow\rangle,\,|t\downarrow\rangle,|b\uparrow\rangle,\,|b\downarrow\rangle$
respectively. $V$ represents the onsite energy term and $T_{x/y}$
represents the hopping term along $x/y$ direction. The effect of the
perpendicular magnetic field $B_z$ is included by adding to the
hopping matrix a phase term $\phi_{\bold x_1,\bold x_2}=\int_{\bold
x_1}^{\bold x_2} \bold A \cdot d\bold l/\phi_0$ where $\bold A=(B_z
y ,0 ,0)$ is the vector potential inducing magnetic field $-B_z$
along $z$ direction and $\phi_0=\hbar/e$. Other parameters in
Hamiltonian (\ref{h2}) are given as : $t_a=\frac{\hbar v_F}{2a}$,
$B_a=B/a^2$, $\tilde m_0=m_0+4B_a$. $a=4nm$ is the lattice constant
along x/y direction. The width of both the central region and six contacts is $480nm$.

In Landauer-B\"uttiker formalism the current flowing out of lead $i$
is given by,\cite{LBformula,long1,sun1}
\begin{eqnarray}\label{LBe}
I_i=\frac{e^2}{h}\sum_jT_{ij}(E_f)(V_i-V_j)
\end{eqnarray}
where $V_i$ is the bias voltage on lead $i$ and
$T_{ij}(E_f)=Tr[\Gamma^iG^r\Gamma^jG^a]$ is the transition
coefficient from lead $j$ to $i$ at Fermi energy $E_f$.
$G^r=[G^a]^\dag=1/(E_f-\tilde H-\sum_{i=1}^6 \Sigma_i^r)$ is the
retarded Green's functions and $\tilde H$ is the Hamiltonian of the
central region. $\Sigma_i^r=[\Sigma_i^a]^\dag$ is the retarded
self-energy coupling to lead $i$ which can be calculated
numerically.\cite{DHLsgf} The linewidth function $\Gamma_i$ is
related to the self-energy by $\Gamma_i=i(\Sigma_i^r-\Sigma_i^a)$.
We apply a small bias $V=V_1-V_4$ across the sample and set leads
$2,3,5,6$ as voltage leads with current being zero. Then from
Eq.(\ref{LBe}), voltages $V_2$, $V_3$, $V_5$ and $V_6$ and the
current $I=I_1=-I_4$ flowing over the sample from lead $1$ to $4$
can be calculated. At last the Hall and longitudinal resistance are
given by $R_H\equiv \frac{V_6-V_2}{I}$ and $R_L\equiv
\frac{V_2-V_3}{I}$ respectively.

\begin{figure*}
\subfigure{\includegraphics[width=1.6\columnwidth,height=0.8\columnwidth]{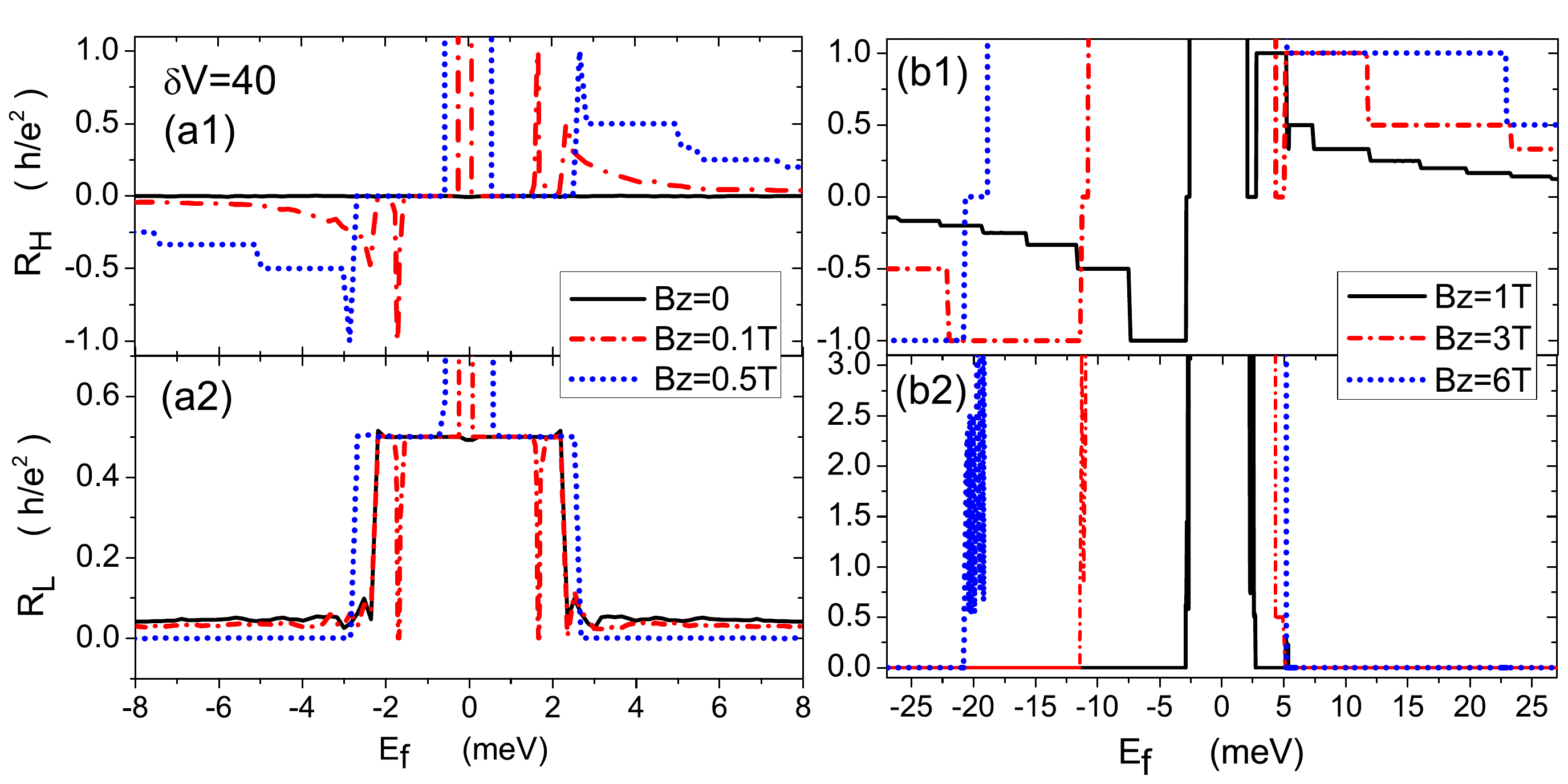}}%
\caption{(Color online) (a,b) Hall resistance $R_H$ and longitudinal resistance $R_L$ as a function of Fermi energy with fixed magnetic field $B_z$. $\delta V=40meV,\,m_0/B<0$.}
  \label{figV40}
\end{figure*}

\begin{figure}
  \centering
\subfigure{
\includegraphics[width=1\columnwidth]{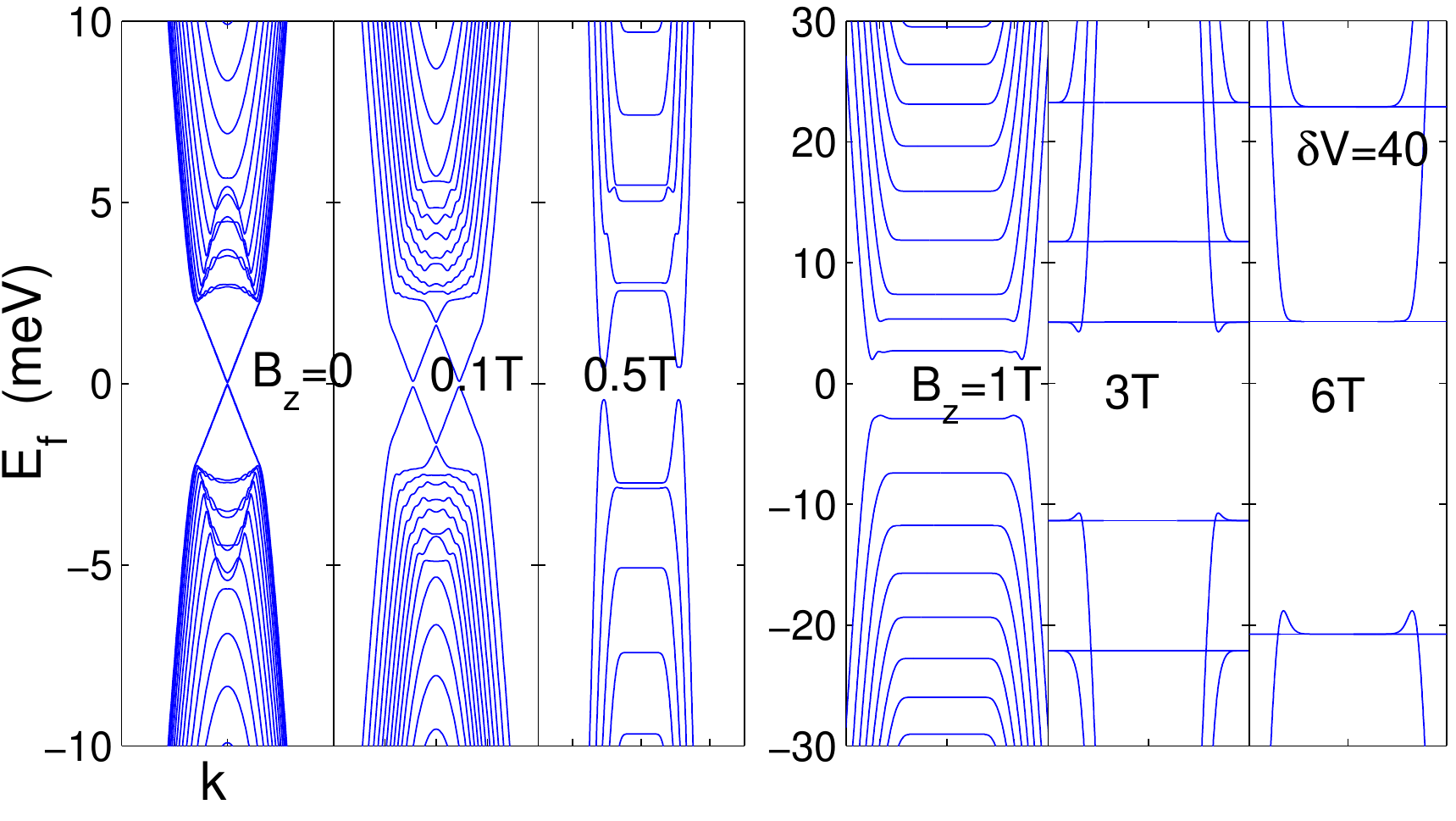}}

\caption{(Color online) The energy dispersion for a nanoribbon with width of 480 nm for several magnetic field. Parameters are identical to those in Fig.\ref{figV40}.
  }
  \label{figebV40}
\end{figure}

The longitudinal and Hall resistances in absence of magnetic field
are shown in Fig.\ref{figB0}, which verify the analysis in
Sec.\ref{sec_phase}. Fig.\ref{figB0}(a1-2) correspond to the
$m_0/B<0$ case. For $\delta V<2\sqrt{M^2-m_0^2}\approx 28meV$, Hall
resistance shows a Hall plateau with the plateau value $h/e^2$ and
the longitudinal resistance is zero at the position of the Hall
plateau. This corresponds to the QAH phase which has been observed
in the recent experiment.\cite{Xue_QAH} For $28mev<\delta
V<\sqrt{M^2-m_0 \hbar^2v_F^2/|B|}\approx 56mev$, the Hall resistance
is almost zero, but the longitudinal resistance shows a plateau with
the plateau value $h/2e^2$, which corresponds to the QPH phase, i.e.
the phase IV in Fig.\ref{figphase}(a1). Increase $\delta V$ anymore, the
resistance at certain regime become infinite corresponding to the
normal-insulator phase VII in Fig.\ref{figphase}(a1). On the other hand, for
$m_0/B>0$ case, the system directly goes from the QAH phase to the
normal-insulator phase without passing the QPH phase by increasing
$\delta V$. Therefore there is no longitudinal resistance plateau, and only the Hall plateau is observed at
$\delta V<2\sqrt{M^2-m_0^2}$ (see Fig.\ref{figB0}(b1-2)).
$R_H$ and $R_L$ are symmetric about $E_f=0$.
The reason is that the Hamiltonian (see Eq.(\ref{h1})) is invariant under the spatial inversion and electron-hole transformation and resultantly the energy band (see Fig.\ref{figphase}(a4,b4)) is symmetric about $E_f=0$.
Magnetic field will destroy electron-hole symmetry and the energy band and resistances will be asymmetric about $E_f=0$ as shown in the following.

In the following we study the effect of magnetic field. We first
consider the case $m_0/B<0$ and $\delta V=0$. The longitudinal and
Hall resistances are shown in Fig.\ref{figV0}. For positive
Fermi energy, there are several integer quantum Hall (IQH) -like
plateaus with zero longitudinal resistance and integer Hall
resistance. These plateaus are consistent with Chern numbers
calculated by Kubo formula and can be explained with LL spectrum
completely. For negative Fermi energy at small magnetic field, see
Fig.\ref{figV0}(a), there is no IQH-like plateau but some exotic
plateaus with nonzero longitudinal resistances. However both
IQH-like and exotic plateaus will form at negative Fermi energy regime for larger magnetic field, see
Fig.\ref{figV0}(b). The exotic plateaus with values
$(R_L,\,R_H)=(\frac{1}{2}h/e^2,0),\,(\frac{2}{9}h/e^2,-\frac{1}{3}h/e^2),\,(\frac{3}{28}h/e^2,-\frac{2}{7}h/e^2)...$,
can be described by
\begin{eqnarray}\label{eqncna}
R_L=\frac{n_a n_c}{n^3_a+n_c^3}\frac{h}{e^2},\,\,R_H=\frac{n_a^2-n_c^2}{n^3_a+n_c^3}\frac{h}{e^2},
\end{eqnarray}
with $(n_a,\,n_c)=(1,1),\,(1,2),\,(1,3)...$ . The first pair is a
QSH-like plateau and the followings are fractional plateaus.
$n_c$ and $n_a$ represent the number of clockwise and anticlockwise
edge states respectively. Suppose that only edge states contribute
to transport without mixing, then Eq.(\ref{eqncna}) can be derived
straightforwardly with Landauer-B\"uttiker formula similar with that
in the QSH effect case.\cite{PRB_HgTe(J.C.Chen)} Therefore these
exotic plateaus originate from the coexistence of clockwise and
anticlockwise edge states.

\begin{figure*}
\subfigure{\includegraphics[width=1.6\columnwidth,height=0.8\columnwidth]{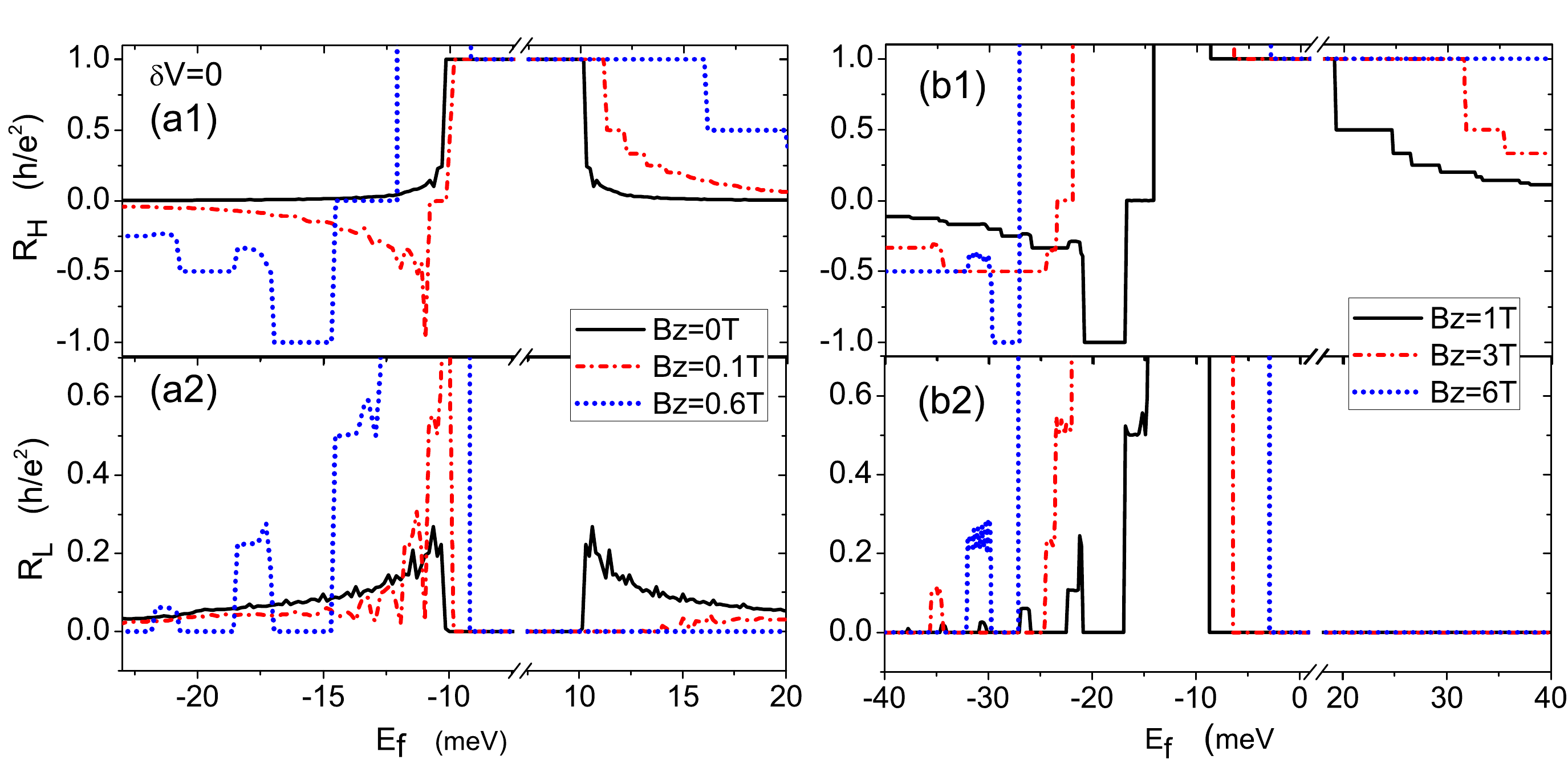}}%
  \caption{(Color online)  (a,b) Hall resistance $R_H$ and longitudinal resistance $R_L$ as a function of Fermi energy with fixed magnetic field $B_z$. $\delta V=0$, $m_0/B>0$.
  }
  \label{figzV0}
\end{figure*}
\begin{figure}
  \centering
\subfigure{
\includegraphics[width=1\columnwidth]{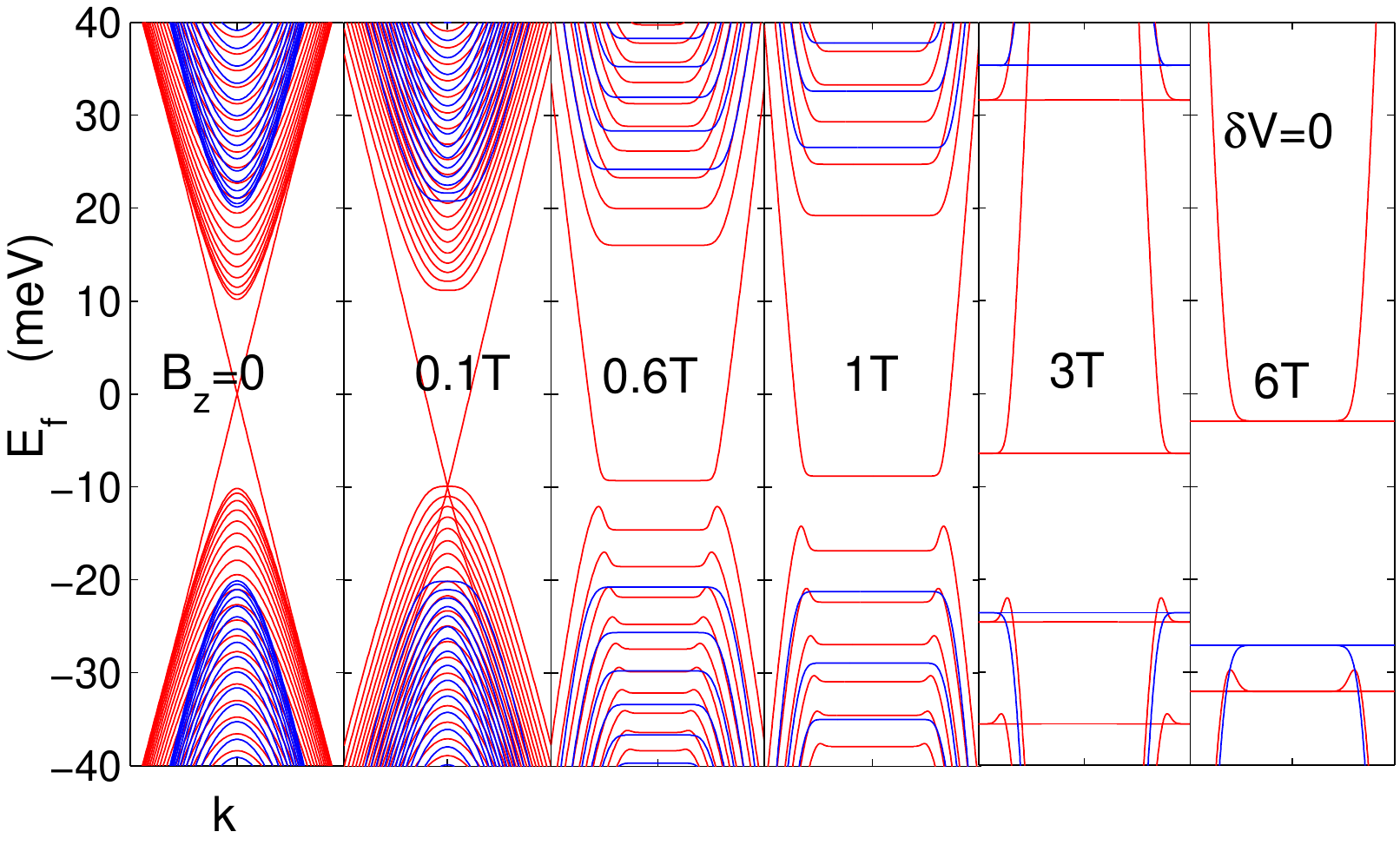}}
\caption{(Color online) The energy dispersion for a nanoribbon with width of 480 nm for several magnetic field. Parameters are identical to those in Fig.\ref{figzV0}.
  }
  \label{figebzV0}
\end{figure}

In order to examine it more clearly, we study the energy dispersion for several magnetic fields as shown in Fig.\ref{figebV0}(a).
Since $\delta V=0$ the Hamiltonian can be decoupled into inverted (see red curves in Fig.\ref{figebV0}(a)) and non-inverted (see blue curves in Fig.\ref{figebV0}(a)) block.
Increasing magnetic field the Dirac cone will move downward into the trivial band.
Then magnetic edge states due to LLs and the intrinsic edge state due to intrinsic energy band will counterpropagate at the same edge of the sample for certain Fermi energy regime.
Consequently a nonzero longitudinal plateau will emerge.
Such analysis is consistent with that by LL spectrum in Fig.\ref{figLLs}(a1).

According to Fig.\ref{figLLs}(a1) the first fractional plateau will only exist below a critical magnetic field about $0.7T$, where the zero mode $E_{0-}$ with positive Chern number cross with the $N=1$ mode of the inner hole-like LLs with negative Chern number.
The critical values for the other fractional plateaus are smaller.
However Fig.\ref{figV0}(b) shows that exotic plateaus can survive even above the critical value but become less perfect.
And they can exist even for $E_f<-20meV$ while the phase with counterpropagate edge states suggested in Fig.\ref{figLLs}(a1) lies between $-20meV<E_f<-10meV$.
The reason is that large enough magnetic field will move the Dirac cone into the nontrivial band.
And these two kinds of edge states will couple to form an exotic energy band with two humps connected to the LL and its edge state, as shown in Fig.\ref{figebV0}(a) and the inset of Fig.\ref{figebV0}(b).
From Fig.\ref{figLLs}(a1), the Chern number in the region $E_f=-21meV$ and $B_z=1T$ (see the green line in Fig.\ref{figebV0}(a) or inset of Fig.\ref{figebV0}(b)) is equal to $-1$, and there should be only a hole edge state (the state C, see inset of Fig.\ref{figebV0}(b)).
However, due to the exotic band with two humps, a pair of extra counterpropagate edge states (states A and B) at one edge of the sample emerge.
Fig.\ref{figebV0}(b) gives the distribution of the wave functions of these extra states A, B and the hole edge state C, and they are indeed edge states.
Therefore, the clockwise and anticlockwise edge states number is $(n_a, n_c)=(1,2)$ to give rise to the first fractional plateau $(R_L\,,R_H)=(\frac{2}{9}h/e^2,-\frac{1}{3}h/e^2)$, although the Chern number is $-1$ as shown in Fig.\ref{figLLs}(a1).

The perfect QSH-like plateau $(R_L,\,R_H)=(\frac{1}{2}h/e^2,\,0)$ is consistent with the QSH-like phase suggested in Fig.\ref{figLLs}(a1) (see the region with Chern number 0).
But unlike the QSH effect, counterpropagating edge states here are caused by magnetic field with time reversal symmetry broken.
The width of this plateau increases to a maximum at $B_z=0.7T$ and then decreases as the magnetic field $B_z$ increases.
It will disappear if the magnetic field is larger than the critical value $B_c=-\frac{\hbar m_0}{e B}\approx4.1T$.

Then we consider the case with SIA potential $\delta V$ but still in the QAH phase with $m_0/B<0,\,\delta V=10meV$.
Fig.\ref{figV10} and Fig.\ref{figebV10} show the longitudinal and Hall resistance and the energy dispersion, respectively.
The inner (outer) bulk bands are nontrivial (trivial).
Moderate magnetic field can move the Dirac cone into the nontrivial band to form the exotic energy band with two humps accompanied by an insulator energy gap.
The resistance will be infinite in the gap region, see Fig.\ref{figV10}.
On the right side of the gap is the QAH plateau and other IQH-like plateaus consistent with the LL spectrum.
On the left side of the gap is the IQH-like and exotic plateaus.
Comparing energy dispersion in Fig.\ref{figebV10} and LL spectrum in Fig.\ref{figLLs}(a2), we find that hole-like bands corresponding to the non-trivial branch form obvious humps while those corresponding to the trivial branch don't.
It indicates that the coupling between intrinsic edge state and trivial band is weaker than that to the nontrivial band,
since the intrinsic edge state is due to the nontrivial band.
Increasing magnetic field, hole-like LLs move downward except $E_{0-}$ (pink line in Fig.\ref{figebV10}) which will move upward below about $B_c= 4.1T$, see Fig.\ref{figLLs}(a2).
For large enough magnetic field, $E_{0-}$ will be the first hole-like LL without humps to vanish the imperfect QSH-like plateau.
But the first fractional plateau will remain since the second hole-like LL corresponds to the nontrivial band, see Fig.\ref{figLLs}(a2).

\begin{figure*}
\subfigure{\includegraphics[width=1.6\columnwidth,height=0.8\columnwidth]{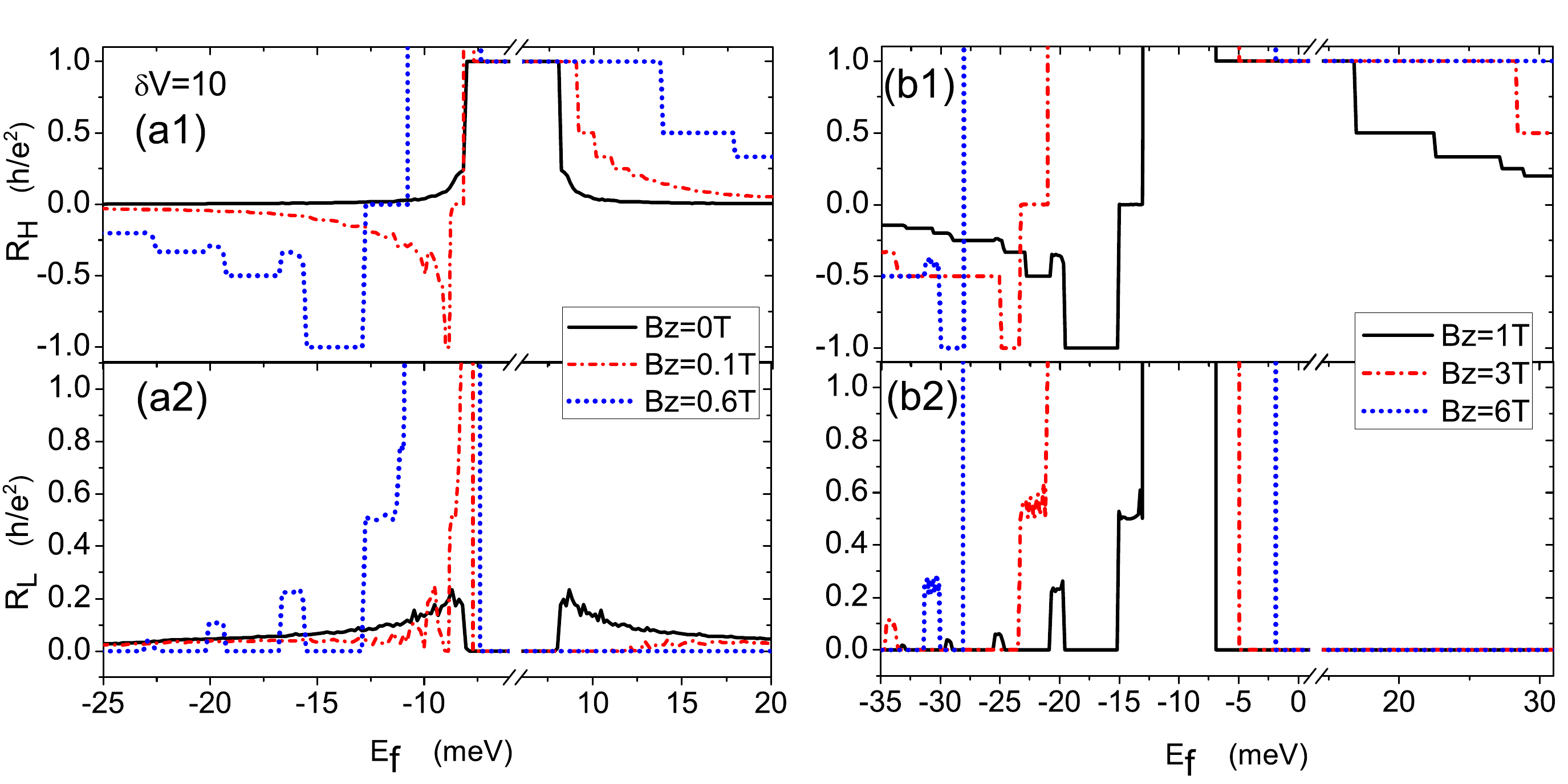}}%
  \caption{(Color online)  (a,b) Hall resistance $R_H$ and longitudinal resistance $R_L$ as a function of Fermi energy with fixed magnetic field $B_z$. $\delta V=10meV$, $m_0/B>0$.
  }
  \label{figzV10}
\end{figure*}

\begin{figure}
  \centering
\subfigure{
\includegraphics[width=1\columnwidth]{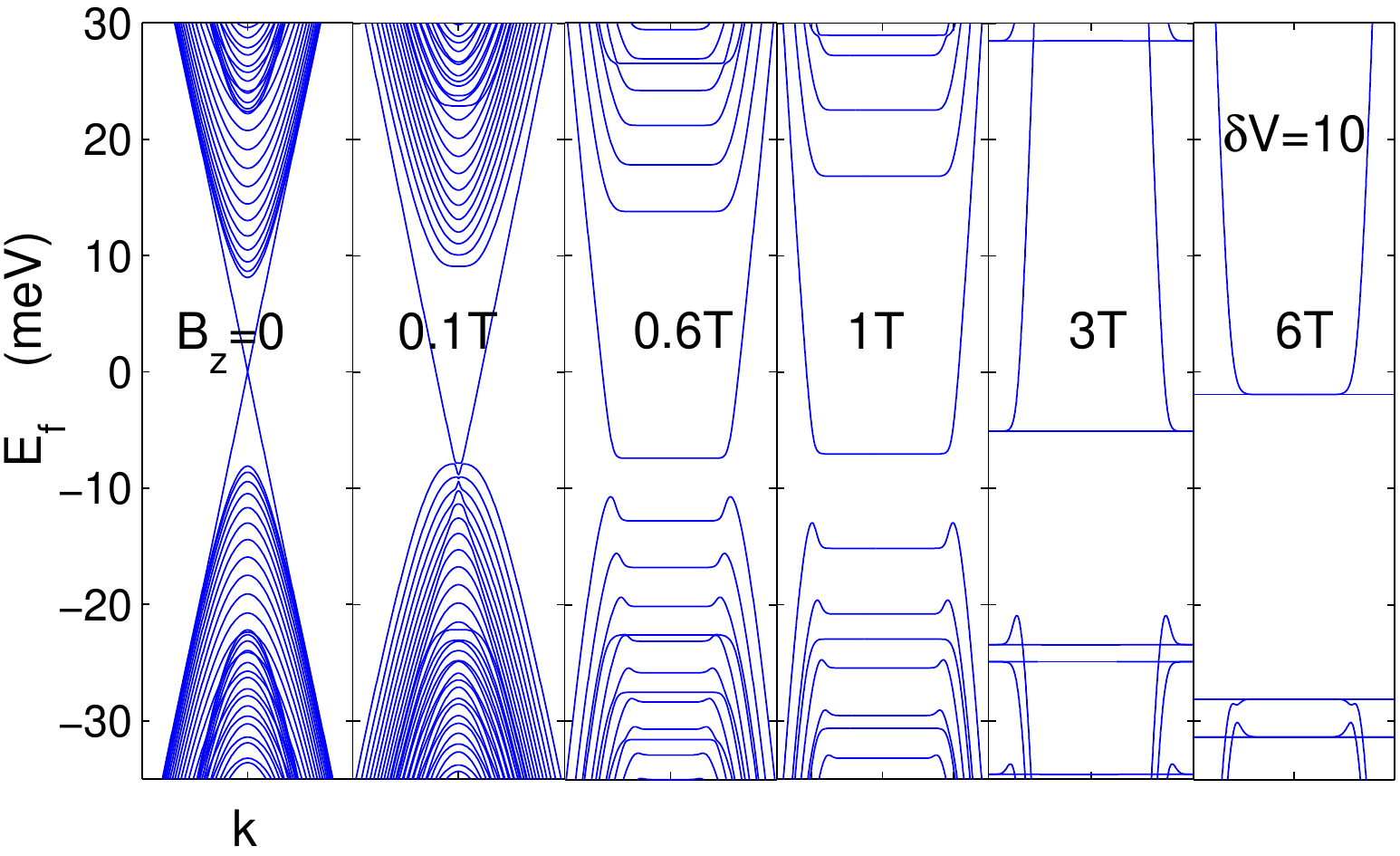}}
\caption{(Color online) The energy dispersion for a nanoribbon with width of 480 nm for several magnetic field. Parameters are identical to those in Fig.\ref{figzV10}.}
  \label{figebzV10}
\end{figure}

In the following, the influence of magnetic field on QPH phase is studied with SIA potential $\delta V=40\,meV$.
Fig.\ref{figV40} show the longitudinal and Hall resistance and Fig.\ref{figebV40} is the energy band.
Magnetic field moves two Dirac cones upward and downward respectively to split the degeneracy.
Therefore there will be exotic plateaus for both negative and positive Fermi energy regime.
But unlike the QAH phase the intrinsic edge state can not enter inside the bulk bands, see as Fig.\ref{figebV10} and Fig.\ref{figebV40}.
Therefore, only the two innermost LL edge states can couple with the intrinsic edge state and form the exotic band with two humps, which cause the QSH-like plateau with the longitudinal resistance $h/2e^2$ and zero Hall resistance at small magnetic field.
The third and following exotic plateaus will not appear.
Increase the magnetic field, two humps gradually fall and the QSH-like plateau disappear. At the large magnetic field, the system come into the IQH regime, in which the quantum Hall plateaus with the plateau value $h/ne^2$ ($n=\pm1,\pm2,...$) well form on both the electron- and hole-sides and the longitudinal resistance is zero, see Fig.\ref{figV40}(b).

Finally, the case $m_0/B>0$ is studied.
Only the QAH phase is considered, since in trivial phase only IQH-like plateaus appear which can be described by LL spectrum completely.
Resistance curves for $\delta V=0$ and $\delta V=10 meV$ are shown in Figs. \ref{figzV0} and \ref{figzV10}.
And they can be understood following the analysis for $m_0/B<0$ with the energy dispersion (Figs.\ref{figebzV0} and \ref{figebzV10}) and LL spectrum (Fig.\ref{figLLs}(b1-2)).
The exotic plateau may be successive for $\delta V=0$, see Fig.\ref{figzV0}(b), but separated by a IQH-like plateau if $\delta V\neq0$, see Fig.\ref{figzV10}(b), similar with that in the case $m_0/B<0$.
But unlike $m_0/B<0$, the inner (outer) bands are nontrivial (trivial) whether SIA potential vanishes or not.
Therefore exotic plateaus are imperfect even for $\delta V=0$.

\begin{figure*}
  \centering
 \hspace{-0.3cm} \subfigure{
    \includegraphics[width=0.75\columnwidth,height=0.7\columnwidth]{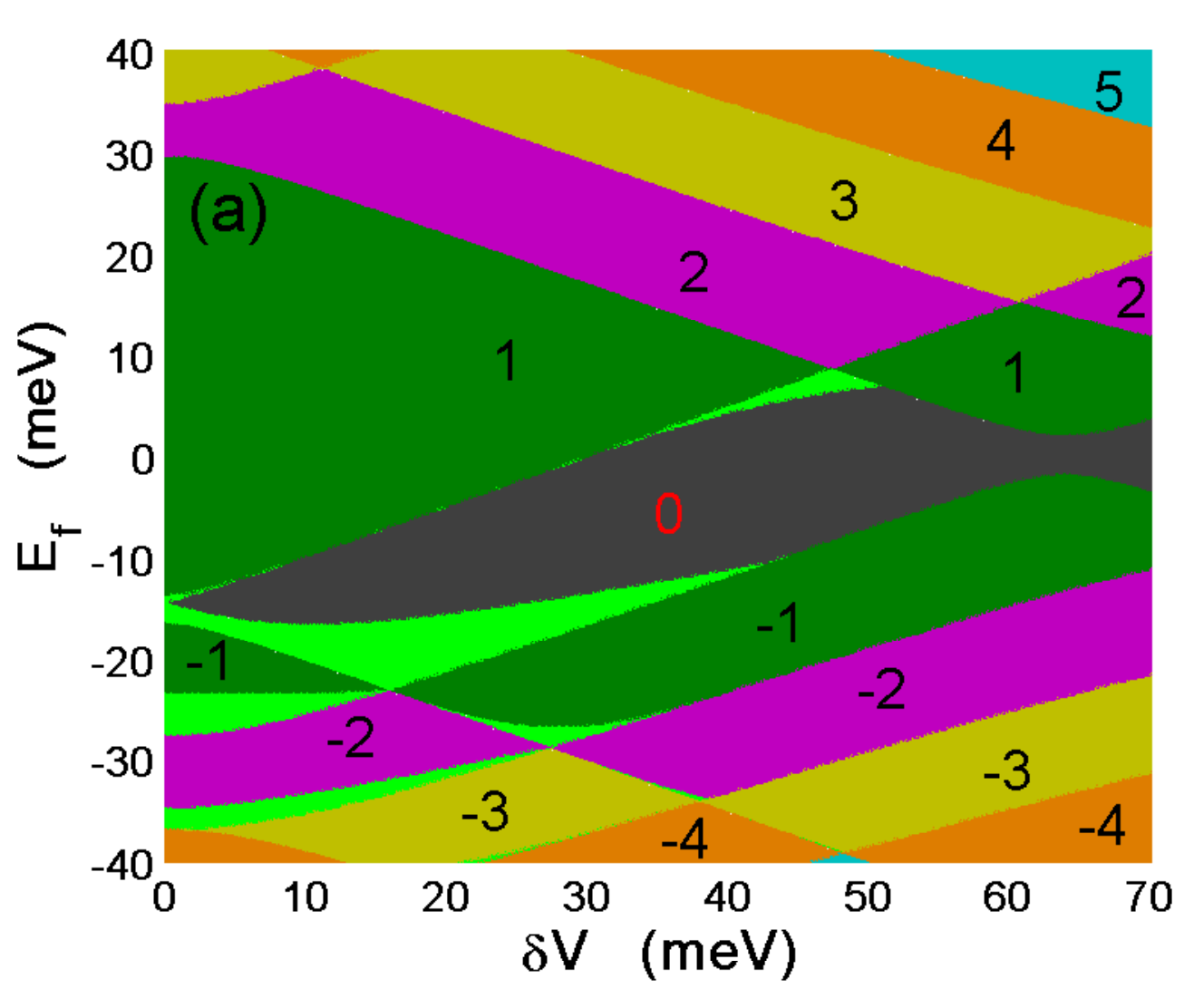}}\hspace{-0.1cm}
  \subfigure{
    \includegraphics[width=0.75\columnwidth,height=0.7\columnwidth]{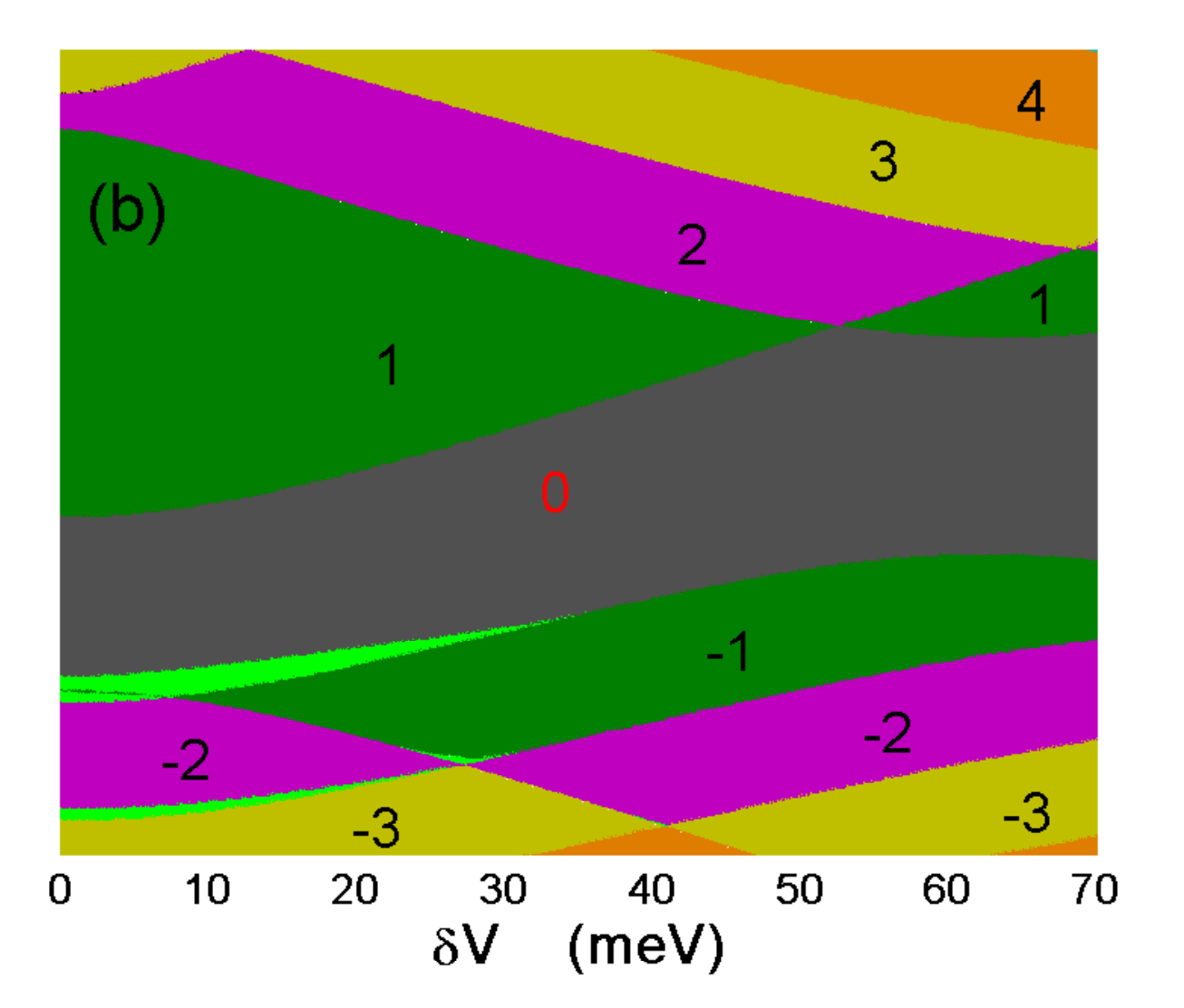}}
  \caption{(Color online) Phase diagram at $B_z=3T$ in the plane of Fermi energy and SIA potential for (a) $m_0/B<0$ and (b) $m_0/B>0$ respectively. Integers show the Hall conductance of the region with zero longitudinal component. Exotic plateaus with nonzero longitudinal component form in the light green region.}
  \label{figB3}
\end{figure*}

To extensively show the effect of SIA potential, the phase diagram at $B_z=3T$ is plotted in Fig.\ref{figB3}.
Integers show the Hall conductance ($G_H=\frac{R_H}{R_H^2+R_L^2}$) of the region with zero longitudinal component ($G_L=\frac{R_L}{R_H^2+R_L^2}$) which are consistent with Fig.\ref{figLLs}(a5,b5).
Light green regions are due to the formation of exotic bands and show the region where the longitudinal resistance doesn't vanish.
They occupy the QH phase region partially and narrow the insulator gap.
The area of light green region is much smaller for $m_0/B>0$ comparing with that for $m_0/B<0$.
For large enough $\delta V$, the light green region will disappear since $\delta V$ leads to the normal-insulator phase.

\section{Conclusion}

In summary we have studied the effect of magnetic field on an ultrathin magnetic topological insulator film with structure inversion asymmetry term.
We give the phase diagram in the plane of the structure-inversion-asymmetry strength and the Fermi energy and derive the Landau level spectrum analytically.
We find that the quantum anomalous Hall phase will survive increasing magnetic field. A new QSH-like phase will form at moderate magnetic field.
Transport properties is simulated in Landauer-B\"uttiker formalism.
Exotic resistance plateau with nonzero longitudinal elements are predicted in a standard Hall bar.
Additionally, the effect of structure inversion asymmetry term on Landau level spectrum and transport behavior is investigated in detail.

\section*{ACKNOWLEDGEMENTS}
This work was financially supported by NBRP
of China (2012CB921303 and 2012CB821402), NSFC of Jiangsu province SBK201340278 and NSF-China under Grants Nos. 11374219, 11274364 and 91221302.


\begin{references}
\bibitem{IQHE}
K.\,v. Klizing, G. Dorda, and M. Pepper, Phys.Rev.Lett {\bf45}, 494 (1980).
\bibitem{Halleffect}
E.\,H. Hall, Am.J.Math. {\bf 2}, 287 (1879).

\bibitem{QAH(Haldane)}
F.\,D.\,M. Haldane, Phys. Rev. Lett. {\bf61},2015 (1988).


\bibitem{Yu(science)}
R. Yu, W. Zhang, H.\,J. Zhang, S.\,C. Zhang, X. Dai, and Z. Fang, Science {\bf329}, 61 (2010).

\bibitem{QAH(proposal)}
M. Onoda and N. Nagaosa, Phys.\,Rev.\,Lett. {\bf90}, 206601 (2003);
X.\,L. Qi, Y.\,S. Wu, and S.\,C. Zhang, Phys.\,Rev.\,B {\bf74}, 085308 (2006);
X.\,L. Qi, T.\,L. Hughes, and S.\,C. Zhang, Phys.Rev.B {\bf78}, 195424 (2008);
C.\,X. Liu, X.\,L. Qi, X. Dai, Z. Fang, and S.\,C. Zhang, Phys.\,Rev.\,Lett. {\bf101}, 146802 (2008);
Z. Qiao, S.\,A. Yang, W.\,X Feng, W.\,K Tse, J. Ding, Y.\,G Yao, J. Wang, and Q. Niu, Phys.Rev.B {\bf82}, 161414 (2010);
K. Nomura, and N. Nagaosa, Phys.Rev.Lett. {\bf106}, 166802 (2011);
H. Jiang, Z.\,H. Qiao, H.\,W. Liu, and Q. Niu, Phys.Rev.B {\bf85}, 045445 (2012);
W.\,K. Tse, Z.\,H. Qiao, Y.\,G. Yao, A.\,H. MacDonald, and Q. Niu, Phys.Rev.B {\bf83}, 155447 (2011);
T.\,W. Chen, Z.\,R. Xiao, D.\,W. Chiou, and G.\,Y. Guo, Phys.Rev.B {\bf84}, 165453 (2011); J. Ding, Z.\,H. Qiao, W.\,X. Feng, Y.\,G. Yao, and Q. Niu, Phys.Rev.B {\bf84}, 195444 (2011).

\bibitem{AHE_E}
E.\,H. Hall, Philos.\,Mag. {\bf12}, 157 (1881);
N. Nagaosa, J. Sinova, S. Onoda, A.\,H. MacDonald, and N.\,P. Ong, Rev. Mod. Phys. {\bf82} 1539 (2010).

\bibitem{QSH_T}
C.\,L. Kane, E.\,J. Mele, Phys. Rev. Lett. {\bf95}, 146802 (2005);
B.\,A. Bernevig and S.\,C. Zhang, Phys. Rev. Lett. {\bf96}, 106802 (2006);
B.\,A. Bernevig, T.\,L. Hughes and S.\,C. Zhang, Science {\bf314}, 1757 (2006).

\bibitem{SpinHall}
M.I. Dyakonov and V.I. Perel, Sov. Phys. JETP Lett. {\bf13}, 467 (1971);
J. E. Hirsch, Phys. Rev. Lett. {\bf83}, 1834 (1999);
Y. K. Kato, R. C. Myers, A. C. Gossard, D. D. Awschalom, Science {\bf306}, 1910 (2004).

\bibitem{QSH_E}
M. K\"onig, S. Wiedmann, C. Br\"une, A. Roth, H. Buhmann, L.\,W. Molenkamp, X.\,L Qi, and S.\,C. Zhang, Science {\bf318}, 766 (2007).

\bibitem{jiang1}
H. Jiang, S. Cheng, Q.-F. Sun, and X. C. Xie, Phys. Rev. Lett. {\bf 103}, 036803 (2009).

\bibitem{Xue_QAH}
C.\,Z. Chang, J.\,S. Zhang, X. Feng, J. Shen, Z.\,C. Zhang, M.\,H. Guo, K. Li, Y.\,B. Ou, P. Wei, L.\,L. Wang, Z.\,Q. Ji, Y. Feng, S.\,H. Ji, X. Chen, J.\,F. Jia, X. Dai, Z. Fang, S.\,C. Zhang, K. He, Y.\,Y. Wang, L. Lu, X.\,C. Ma, and Q.\,K. Xue, Science {\bf340}, 167 (2013).

\bibitem{review(Ti)}
M. Z. Hasan and C. L. Kane, Rev. Mod. Phys. {\bf82}, 3045 (2010);
X.\,L. Qi, S.\,C. Zhang, Rev. Mod. Phys. {\bf83}, 1057 (2011).

\bibitem{NJP(Shan)}
W.\,Y. Shan, H.\,Z. Lu, and S.\,Q. Shen, New J. Phys. \textbf{12}, 043048 (2010).
\bibitem{PRB(Lu)}
H.\,Z. Lu, W.\,Y. Shan, W. Yao, Q. Niu, and S.\,Q. Shen Phys. Rev. B \textbf{81}, 115407 (2010);
H.\,C. Li, L. Sheng, D.\,N. Sheng, and D.\,Y. Xing Phys. Rev. B \textbf{82}, 165104 (2010).

\bibitem{LLTIfilm}
A.\,A. Zyuzin and A.\,A. Burkov, Phys. Rev. B {\bf83} 195413 (2011).
\bibitem{PRB(L.Sheng)}
H.\,C. Li, L. Sheng, and D.\,Y. Xing Phys. Rev. B \textbf{84}, 035310 (2011).

\bibitem{effective_H_Liu}
C.\,X. Liu, X.\,L. Qi, H.\,J. Zhang, X. Dai, Z. Fang, and S.\,C. Zhang, Phys. Rev. B {\bf82}, 045122 (2010).

\bibitem{JAPLL}
M. Tahir, K. Sabeeh, and U. Schwingenschl\"ogl, J. Appl. Phys. {\bf113}, 043720 (2013).

\bibitem{book_QED1965}
A. I. Akheizer and V. B. Berestetsky, $Quantum$ $Electrodynamics$ (Interscience, New York, 1965).

\bibitem{DiracLL_TI_E}
P. Cheng, C.\,l. Song, T. Zhang, Y.\,Y. Zhang, Y.\,L. Wang, J.\,F. Jia, J. Wang, Y.\,Y. Wang, B.\,F. Zhu, X. Chen, X.\,C. Ma, K. He, L.\,L. Wang, X. Dai, Z. Fang, X.\,C. Xie, X.\,L. Qi, C.\,X. Liu, S.\,C. Zhang, and Q.\,K. Xue, Phys.Rev.Lett {\bf105}, 076801 (2010);
T. Hanaguri, K. Igarashi, M. Kawamura, H. Takagi, and T. Sasagawa, Phys. Rev. B {\bf82}, 081305 (2010);
B. Sacepe, J. B. Oostinga, J. Li, A. Ubaldini, N. J. G. Couto, E. Giannini, and A. F. Morpurgo, Nature Commun. {\bf2}, 575 (2011).
\bibitem{PRL(Y.P.Jiang)}
Y.\,P Jiang, Y.\,L. Wang, M. Chen, Z. Li, C.\,L. Song, K. He, L.\,L. Wang, X. Chen, X.\,C. Ma, and Q.\,K. Xue, Phys. Rev. Lett {\bf108}, 016401 (2012).
\bibitem{DLL(Okada)}
Y. Okada, W.\,W. Zhou, C. Dhital, D. Walkup, Y. Ran, Z. Wang, S.\,D. Wilson, and V. Madhavan, Phys. Rev. Lett {\bf109}, 166407 (2012).
\bibitem{DLL(HgTe)}
C. Br\"une, C.\,X. Liu, E.\,G. Novik, E.\,M. Hankiewicz, H. Buhmann, Y.\,L. Chen, X.\,L. Qi, Z.\,X. Shen, S.\,C. Zhang, and L.\,W. Molenkamp, Phys.Rev.Lett {\bf106}, 126803 (2011).

\bibitem{lu2013QAH}
H.\,Z Lu, A. Zhao, and S.\,Q Shen, Phys. Rev. Lett {\bf111}, 146802 (2013).

\bibitem{HJZhang}
H.\,J. Zhang, C.\,X. Liu, X.\,L. Qi, X. Dai, Z. Fang, and S.\,C. Zhang, Nat. Phys. {\bf5}, 438 (2009).

\bibitem{note1}
Parameters given as:
$\tilde A_2=-iA_2\langle\psi(A_1)|\sigma_x|\chi(-A_1)\rangle$, $B=(\tilde B_2-\tilde B_1)/2$, $\tilde B_1=B_2\langle\psi(A_1)|\sigma_z|\psi(A_1)\rangle$, $\tilde B_2=B_2\langle\chi(A_1)|\sigma_z|\chi(A_1)\rangle$ and $m_0=(E_+-E_-)/2$.
$\psi(A_1)$ and $\chi(A_1)$ are the eigenvectors of Hamiltonian $h(A_1)=\left(\begin{array}
  {cc}-(D-B_1)\partial_z^2+C+M&-iA_1\partial_z\\-iA_1\partial_z&-(D+B_1)\partial_z^2+C-M
\end{array}\right)$ at $\Gamma$ point, and $E_{\pm}$ are the corresponding eigenvalues. See Ref.(\mycite{NJP(Shan)}) for detail of derivation.

\bibitem{QSH_NI_Osci(Liu)}
C.\,X. Liu, H.\,J. Zhang, B.\,H. Yan, X.\,L. Qi, T. Frauenheim, X. Dai, Z. Fang, and S.\,C. Zhang, Phys. Rev. B {\bf81}, 041307(R) (2010).

\bibitem{PRB(L.Sheng2012)}
H.\,C. Li, L. Sheng, and D.\,Y. Xing, Phys. Rev. B \textbf{85}, 045118 (2012).

\bibitem{nphys(Y.Zhang)}
Y. Zhang, K. He, C.-Z. Chang, C.-L. Song, L.-L. Wang, X. Chen,
J.-F. Jia, Z. Fang, X. Dai, W.-Y. Shan, S.\,-Q. Shen, Q. Niu, X.\,-L. Qi, S.\,-C. Zhang, X.\,-C. Ma, and Q.\,-K Xue, Nat. Phys. \textbf{6}, 584 (2010).

\bibitem{PRB_HgTe(J.C.Chen)}
J.\,C. Chen, J. Wang, and Q.\,F. Sun, Phys. Rev. B \textbf{85}, 125401 (2012).
\bibitem{jiang2}
H. Jiang, L. Wang, Q.-F. Sun, and X. C. Xie, Phys. Rev. B {\bf 80}, 165316 (2009).

\bibitem{KuboChern}
D.\,J. Thouless, M. Kohmoto, M.\,P. Nightingale, and M.\,D. Nijs, Phys. Rev. Lett. {\bf49}, 405 (1982);
M. Kohmoto, Ann. Phys. (NY) {\bf160}, 343 (1985).

\bibitem{note2}
Chern number is calculated directly with Eq.(\ref{ChernNumber}) in the discrete k-space
with discretized Hamiltonian given in Eq.(\ref{h2}).
We find it is accurate enough to discretize the first Brillouin zone up to $201\times201$ lattice points.
For each k mesh the Hamiltonian is $4N_y \times 4 N_y$.
$N_y=1$ in the absence of magnetic field and
$N_y$ satisfies $N_y\phi=2\pi$ and $\phi=B_z a^2$ with the presence of magnetic field.

\bibitem{PRL(2004)(S.Q.Shen)(SR)}
S.\,Q. Shen, M. Ma, X.\,C. Xie, and F.\,C. Zhang, Phys. Rev. Lett {\bf92}, 256603 (2004).

\bibitem{grapheneLL}
V.\,P. Gusynin and S.\,G. Sharapov, Phys. Rev. Lett {\bf95}, 146801 (2005).

\bibitem{DiracLL_graphene_E}
K. S. Novoselov, A. K. Geim, S. V. Morozov, D. Jiang, M. I. Katsnelson, I. V. Grigorieva, S. V. Dubonos, and A. A. Firsov, Nature (London) {\bf438}, 197 (2005).
Y. Zhang, Y.-W. Tan, H. L. Stormer, and P. Kim, Nature (London) {\bf438}, 201 (2005).

\bibitem{LBformula}
S. Datta, $Electronic$ $Transport$ $in$ $Mesoscopic$ $Systems$ ( Cambridge University Press, Cambridge, England, 1995 ).

\bibitem{long1}
W. Long, Q.-F. Sun, and J. Wang, Phys. Rev. Lett. {\bf 101}, 166806 (2008).

\bibitem{sun1}
Q.-F. Sun and X. C. Xie, Phys. Rev. Lett. {\bf 104}, 066805 (2010).

\bibitem{DHLsgf}
D.\,H. Lee and J.\,D. Joannopoulos, Phys. Rev. B \textbf{23}, 4997 (1981);
M.\,P.\,L. Sancho, J.\,M.\,L. Sancho, and J. Rubio, J. Phys. F {\bf15}, 851 (1985).


\end{references}
\end{document}